\documentclass[11pt,a4paper]{article}
\usepackage{jcappub}

\usepackage{comment} 
\usepackage{slashed}

\newcommand{\lsim}{\mathrel{\mathop{\kern 0pt \rlap
  {\raise.2ex\hbox{$<$}}}
  \lower.9ex\hbox{\kern-.190em $\sim$}}}
\newcommand{\gsim}{\mathrel{\mathop{\kern 0pt \rlap
  {\raise.2ex\hbox{$>$}}}
  \lower.9ex\hbox{\kern-.190em $\sim$}}}

\begin{document}



\title{Constraining dark matter using 20-year INTEGRAL/IBIS \\ observations II: Sub-GeV dark matter}

\author[a]{Jordan Koechler}
\author[b,c]{and Pedro De la Torre Luque}

\affiliation[a]{Istituto Nazionale di Fisica Nucleare, Sezione di Torino, \\ Via P. Giuria 1, 10125 Torino, Italy}
\affiliation[b]{Departamento de F\'isica Te\'orica, \\ M-15, Universidad Aut\'onoma de Madrid, E-28049 Madrid, Spain}
\affiliation[c]{Instituto de F\'isica Te\'orica UAM-CSIC, \\ Universidad Aut\'onoma de Madrid, C/ Nicol\'as Cabrera, 13-15, 28049 Madrid, Spain}

\emailAdd{jordan.koechler@gmail.com}
\emailAdd{pedro.delatorre@uam.es}

\abstract{
We present new constraints on sub-GeV dark matter (DM) annihilation and decay using twenty years of diffuse hard X-ray observations from the INTEGRAL/IBIS telescope. By performing a joint analysis of the spectral and morphological information contained in the IBIS dataset, we simultaneously fit Galactic longitudinal profiles and energy spectra, exploiting their complementary signatures to disentangle potential DM signals from the dominant astrophysical backgrounds.
We use physically motivated background models combining inverse-Compton (IC) emission from Galactic cosmic-ray electrons with thermal emission from unresolved accreting white dwarfs, and consistently include the secondary IC emission generated by DM-induced electron and positron populations propagated through the Galaxy. 
We conservatively derive 95\% upper limits on the annihilation cross section $\langle\sigma v\rangle$ and decay half-life $\tau$ for the $e^+e^-$, $\mu^+\mu^-$, and $\pi^+\pi^-$ final states across the mass range $1$\,MeV to $5$\,GeV. Our constraints are in the range $ 10^{-28}$ cm$^3$/s $\lesssim \langle\sigma v\rangle \lesssim 10^{-26}$ cm$^3$/s for annihilation and $\tau \gtrsim 10^{25}$ s for decay at DM masses of a few MeV, improving upon existing X-ray limits by up to two orders of magnitude over much of the mass range. 
Our bounds are translated into constraints on the kinetic mixing parameter $\epsilon$ of a dark photon  and the scalar-Higgs mixing angle $\sin\theta$, and compared with existing indirect, direct and accelerator limits available, providing a comprehensive picture of the allowed parameter space for light portal DM.
}


\arxivnumber{2609.xxxxx}

\maketitle

\section{Introduction}

Light dark matter (DM) is a theoretically well-motivated target for indirect searches: it arises naturally in extensions of the Standard Model (SM) containing new light degrees of freedom, and its interactions with the SM can remain sufficiently weak to evade conventional direct and collider probes. Vector and scalar portals provide simple, renormalizable realizations of such scenarios, and they give rise to distinctive relations between annihilation, decay, and scattering rates that allow limits derived from one observable to be translated into constraints on fundamental couplings. Identifying the most sensitive channels for such searches, and understanding what ultimately limits their reach, is therefore a central task for sub-GeV DM phenomenology.

X-ray observations have emerged as one of the most powerful probes of sub-GeV DM, with recent analyses demonstrating competitive and, in some cases, leading constraints on both annihilation and decay rates~\cite{Cirelli:2020bpc,Cirelli:2023tnx,DelaTorreLuque:2023olp,Balaji:2025afr}. This sensitivity arises because sub-GeV DM generically produces photons in the X-ray band, either through direct emission or via secondary processes. In particular, DM annihilation and decay into electrons and positrons leads to inverse-Compton (IC) scattering off ambient radiation fields, producing a diffuse, extended X-ray signal whose spectral and spatial properties are directly tied to the underlying particle-physics model. For portal scenarios, where the DM couples to the SM through a light mediator, these secondary IC signatures can dominate the observable signal over a broad range of DM masses, and their morphology encodes information about both the DM distribution and the propagation environment.

The robustness and ultimate reach of these constraints, however, critically depend on how accurately the X-ray background can be modelled. In the hard X-ray band, Galactic diffuse emission is expected to arise primarily from two components: IC scattering of cosmic-ray (CR) electrons on interstellar radiation fields (ISRFs), which dominates above several tens of keV, and thermal emission from unresolved compact astrophysical sources, including accreting white dwarfs and other stellar remnants~\cite{Krivonos:2006px}. Both components carry substantial uncertainties — in the CR electron spectrum, the ISRF intensity, and the population properties of unresolved sources — and current data do not yet permit a fully consistent, morphologically constrained description of all contributions. Reducing these uncertainties is therefore a prerequisite for extracting reliable DM limits, and forms a central goal of this work.

The Imager on Board the INTEGRAL Satellite (IBIS), onboard the INTEGRAL observatory, provides a unique dataset for this purpose~\cite{Ubertini:2003ih}. INTEGRAL (INTErnational Gamma-Ray Astrophysics Laboratory) was a space-based mission designed to observe the sky in hard X-rays and soft $\gamma$-rays, combining imaging, spectroscopy, and timing capabilities over a broad energy range. Its IBIS instrument, optimized for high-angular-resolution imaging in the $\sim$15\,keV to several MeV range, employs a coded-mask technique together with layered detector systems to reconstruct both the spectral and spatial distribution of incoming radiation. This design enables IBIS to disentangle diffuse emission from point sources and to extract detailed morphological information across the sky. Recent analyses~\cite{Krivonos:2020qvl} have exploited these capabilities to characterize the spatial and spectral properties of the Galactic diffuse hard X-ray emission, providing the observational foundation on which the present work builds.

In this work, we extend the analysis framework developed in our companion study of primordial black hole evaporation~\cite{Koechler:2026dci} to the case of particle DM in vector and scalar portal scenarios. We exploit the full constraining power of the IBIS observations by performing a joint analysis of their spectral and morphological information, searching for additional diffuse emission components associated with DM annihilation and decay. We construct a physically motivated model for the Galactic X-ray background--combining CR electron IC emission with unresolved thermal emission--and consistently include the secondary IC contribution from DM-induced electron and positron populations, whose spectral and spatial properties are determined by the portal model. By simultaneously fitting the spatial and spectral data, we disentangle potential DM signals from astrophysical backgrounds more effectively than spectral analyses alone allow. 
This approach substantially tightens existing bounds over the mass range where secondary IC emission dominates, while delivering a more robust 
 characterisation of the diffuse hard X-ray sky.

\section{INTEGRAL/IBIS observations}
\label{sec:data}

In this work, we use the IBIS/ISGRI dataset released by Ref.~\cite{Krivonos:2024bih}, comprising longitudinal emission profiles and energy spectra of the diffuse Galactic hard X-ray emission extracted from observations spanning May 2003 to February 2024 (INTEGRAL revolutions 70--2740). The data cover three broad energy bands --- 25--60, 60--80, and 80--200 keV --- and we extracted the observations in three extended sky regions: a $20^\circ\times20^\circ$ Galactic Bulge window (\texttt{GB20}) centred on $(\ell,b)=(0^\circ,0^\circ)$, and two high-longitude windows of $30^\circ\times20^\circ$ centred at $\ell=+80^\circ$ (\texttt{L+80}) and $\ell=-80^\circ$ (\texttt{L-80}). Together, the three longitudinal profiles (extracted within $|b|<10^\circ$) and three regional energy spectra yield 126 flux measurements across 21 logarithmically spaced bins between 25 and 185 keV. These form the basis of our joint spectral and morphological analysis. For full details of the data reduction and extraction procedure, we refer the reader to our companion paper~\cite{Koechler:2026dci}; here we summarize only the features directly relevant to the present analysis.

The IBIS instrument employs a coded-mask design that combines a large field of view ($\sim29^\circ\times29^\circ$, fully coded to $9^\circ\times9^\circ$) with an angular resolution of $\sim12'$ FWHM. This allows resolved point sources to be reconstructed and subtracted, while the wide field of view enables the instrument to be operated as a collimated detector sensitive to large-scale diffuse emission. The Galactic signal is inferred directly from the detector count rate, after removal of point sources and instrumental background, rather than from sky images --- a strategy that reduces systematic uncertainties and is well suited to recovering the morphology of extended structures on scales comparable to the Galactic bulge and disk.

A critical requirement for this approach is an accurate background model, since the instrumental background dominates the total count rate. Ref.~\cite{Krivonos:2024bih} introduces a continuously calibrated background model in which the detector sensitivity evolution over the mission lifetime is tracked through Crab Nebula observations, and high-latitude pointings are used to estimate and subtract the isotropic components. The resulting residual emission traces the Galactic hard X-ray background with well-controlled systematics across the full 25--200 keV range.

The Galactic background is significantly detected up to $\sim200$\,keV. At low energies ($\lesssim60$\,keV), the emission morphology closely follows the stellar mass distribution traced by near-infrared observations~\cite{Krivonos:2024bih}, consistent with a dominant contribution from unresolved compact sources in the Galactic Ridge. At higher energies, the emission becomes more extended and is increasingly dominated by CR electron IC scattering. This energy-dependent morphological variation is precisely what makes the joint spectral and morphological dataset particularly powerful for disentangling astrophysical foregrounds from a potential dark matter signal.

Because the 126 flux measurements share common INTEGRAL revolutions and a common background-modeling procedure, they are not statistically independent. We account for this by estimating the full $126\times126$ covariance matrix using a non-parametric cluster bootstrap, taking the INTEGRAL revolution as the resampling unit and applying Ledoit--Wolf shrinkage~\cite{LEDOIT2004365} to regularize the empirical covariance. Positive definiteness is verified via Cholesky decomposition before the matrix enters the likelihood. All the details are given in Ref.~\cite{Koechler:2026dci}.

\section{Hard $X$-rays and soft $\gamma$-rays from light dark matter}

In this section, we discuss the processes that generate hard X-ray and soft $\gamma$-ray emission from sub-GeV DM annihilation and decay. In this mass range, the kinematically accessible final states are $\mathrm{DM}\,(\mathrm{DM}) \to e^+e^-,\,\mu^+\mu^-,\,\pi^+\pi^-$, which open when $m_{\rm DM} > m_i$ for annihilation and $m_{\rm DM} > 2m_i$ for decay, where $i \in \{e,\mu,\pi\}$.

For a given channel, the total photon flux receives two contributions: (i) \emph{prompt emission}, arising from final-state radiation off charged particles and from radiative decays of heavier final states (in particular muons and pions); and (ii) \emph{inverse-Compton (IC) emission}, produced when DM-induced $e^\pm$ upscatter ambient Galactic photons. Which contribution dominates depends on the DM mass and the final state: for $e^+e^-$ at low masses the prompt component is suppressed and IC dominates, while for heavier leptons prompt emission becomes increasingly important.

\subsection*{Prompt emission}

The prompt photon flux is
\begin{equation}
    \frac{d\Phi_{\gamma,\rm prompt}}{dE_\gamma\, d\Omega}
    = \frac{1}{4\pi}\,\frac{dN_{\gamma,\rm prompt}}{dE_\gamma}
    \times
    \begin{cases}
        \dfrac{\langle\sigma v\rangle}{2\xi m_{\rm DM}^2}\,\mathcal{J}(\ell,b) & \textrm{(annihilation)} \\[10pt]
        \dfrac{\Gamma}{m_{\rm DM}}\,\mathcal{D}(\ell,b) & \textrm{(decay),}
    \end{cases}
    \label{eq:prompt_flux}
\end{equation}
where $dN_{\gamma,\rm prompt}/dE_\gamma$ is the prompt photon spectrum per annihilation or decay, computed following Ref.~\cite{Cirelli:2020bpc}, and $\xi = 1,2$ for bosonic and fermionic DM respectively. The $J$- and $D$-factors,
\begin{equation}
    \mathcal{J}(\ell,b) = \int_{\rm l.o.s.} ds\,\rho_{\rm DM}^2\!\left(r(s,\ell,b)\right), \qquad
    \mathcal{D}(\ell,b) = \int_{\rm l.o.s.} ds\,\rho_{\rm DM}\!\left(r(s,\ell,b)\right),
\end{equation}
are line-of-sight integrals of the DM density profile to the second and first power, respectively, with $(\ell,b)$ denoting Galactic longitude and latitude. We adopt a Navarro--Frenk--White (NFW) profile~\cite{Navarro:1995iw} with scale radius $r_s = 20$\,kpc and local density $\rho_{\rm DM}(r_\odot = 8.5\,\mathrm{kpc}) = 0.4$\,GeV\,cm$^{-3}$.

\subsection*{Inverse-Compton emission}

For the IC contribution, we follow the same procedure as in Ref.~\cite{Koechler:2026dci}: the $e^\pm$ injected by DM annihilation or decay are propagated through the Galaxy, and their interactions with the Galactic interstellar radiation field (ISRF) are used to compute the resulting X-ray emission. The $e^\pm$ source term is
\begin{equation}
    Q_e(r,E_e) =
    \begin{cases}
        \dfrac{\langle\sigma v\rangle}{2\xi}
        \left(\dfrac{\rho_{\rm DM}(r)}{m_{\rm DM}}\right)^{\!2}
        \dfrac{dN_e^{\rm ann}}{dE_e} & \textrm{(annihilation)} \\[12pt]
        \Gamma\,
        \left(\dfrac{\rho_{\rm DM}(r)}{m_{\rm DM}}\right)
        \dfrac{dN_e^{\rm dec}}{dE_e} & \textrm{(decay),}
    \end{cases}
    \label{eq:source_term}
\end{equation}
where $dN_e/dE_e$ is the $e^\pm$ energy spectrum per annihilation or decay. Unlike in our companion paper~\cite{Koechler:2026dci}, which considers PBH evaporation, here this spectrum is determined entirely by the DM particle-physics model and the choice of final state.

The injected $e^\pm$ spectra are computed as follows. For $e^+e^-$ final states, the spectrum is a monochromatic line at $E_e = m_{\rm DM}$ (annihilation) or $E_e = m_{\rm DM}/2$ (decay). For $\mu^+\mu^-$ final states, the muons decay via $\mu \to e\,\nu_e\,\nu_\mu$, producing a Michel spectrum~\cite{Michel:1950} boosted to the DM rest frame~\cite{Donato:2003xg}. For $\pi^+\pi^-$ final states, the $e^\pm$ arise through the full $\pi^\pm$ decay chain ($\pi\to\mu\to e$) and are computed accordingly. We have verified that the bremsstrahlung contribution to the photon flux is negligible across our energy range: the bremsstrahlung cross-section scales with electron energy and peaks near 1\,GeV, making it subdominant well below the MeV scale relevant here.

The propagation of the injected $e^\pm$ through the Galaxy is handled numerically using \texttt{DRAGON2}~\cite{DRAGON2-1,DRAGON2-2}, which solves the cosmic-ray transport equation including diffusion, energy losses, and convection. The resulting propagated $e^\pm$ density $dn_e/dE_e$ is then passed to \texttt{HERMES}~\cite{Dundovic:2021ryb}, which computes the IC emissivity and integrates it along the line of sight to produce the IC photon flux $d\Phi_{\gamma,\rm IC}/(dE_\gamma\,d\Omega)$ as a HEALPix sky map. From this map, we extract the predicted flux in the same sky regions, longitude bins, and energy bands defined in Section~\ref{sec:data}, expressed in Crab units using the Crab spectrum $d\Phi_\gamma^{\rm Crab}/dE_\gamma = 10\,E_\gamma^{-2.1}$\,cm$^{-2}$\,s$^{-1}$\,keV$^{-1}$~\cite{Churazov:2006bk}.

\section{Background modelling and analysis procedure}
\label{sec:analysis}

The Galactic hard X-ray sky up to MeV energies is dominated by two astrophysical backgrounds: IC emission from CR electrons above a few tens of keV, and thermal emission from unresolved accreting white dwarfs (WDs) at lower energies. We refer the reader to our companion paper~\cite{Koechler:2026dci} for a detailed discussion of their physical origin and observational evidence; here we focus on how both components are modelled and fitted to the IBIS data to extract constraints on DM annihilation and decay.

\subsection{Astrophysical background model}
\label{subsec:backgrounds}

\paragraph{CR electrons.} The diffuse X-ray and soft $\gamma$-ray emission from CR electrons is modelled using the propagation framework of Refs.~\cite{delaTorreLuque:2022vhm, DelaTorreLuque:2023zyd}. A key feature of this model is that it is constrained not only by local CR measurements but also by the large-scale diffuse $\gamma$-ray emissivity observed by Fermi-LAT down to tens of MeV~\cite{Casandjian:2015hja}, making it well suited for predicting the Galactic-scale X-ray emission probed by IBIS. IC emission is evaluated using the three-dimensional ISRF models of Ref.~\cite{2016PhRvD..94f3009V}; bremsstrahlung, which is subdominant across most of the IBIS energy range, is computed from detailed spiral-arm gas maps. The model has been validated against a broad set of X-ray and $\gamma$-ray observations~\cite{laTorreLuquePedro:2024est, DelaTorreLuque:2024zsr} and, as shown below, reproduces the IBIS data above $\sim60$\,keV without any tuning. Its overall normalization $A_{\rm CR}$ is left free in the fit to absorb uncertainties in the CR electron distribution and ISRF, while the spectral shape and spatial morphology are held fixed.

\paragraph{Unresolved white dwarfs.} The low-energy ($\lesssim60$\,keV) background is dominated by unresolved accreting WDs, whose cumulative emission traces the stellar mass distribution of the Galaxy — closely following the Galactic NIR brightness — and whose spectral shape is well described by emission from intermediate polars (IPs), with a characteristic minimum near $\sim80$\,keV~\cite{Krivonos:2006px, Krivonos:2024bih}. For the spectrum, we adopt the IP template of Ref.~\cite{Suleimanov:2004aq}, parameterized by an average WD mass $M_{\rm WD} \in [0.3, 1.4]\,M_\odot$ — with previous measurements favouring values in the range $0.5$--$0.66\,M_\odot$~\cite{Krivonos:2006px, Turler:2010pm, Yuasa:2012qe} — and a normalization $A_{\rm IP}$. For the morphology, we use the COBE/DIRBE NIR profile at $4.9\,\mu$m convolved with the IBIS/ISGRI collimator response~\cite{Krivonos:2024bih}, with a normalization $A_{\rm NIR}^{\Delta E}$ free in each energy band.

The spectral and spatial descriptions are coupled by requiring consistency between the band-averaged IP flux and the region-averaged NIR profile for each sky region and energy band:
\begin{equation}
    \frac{A_{\rm IP}^{\rm reg}}{\Delta E} \int_{\Delta E} dE\, \Phi_{\rm IP}(E; M_{\rm WD}^{\rm reg})
    = \frac{A_{\rm NIR}^{\Delta E}}{\Delta\ell_{\rm reg}} \int_{\Delta\ell_{\rm reg}} d\ell\, \Phi_{\rm NIR}^{\Delta E}(\ell)\;.
    \label{eq:NIR_IP_consistency}
\end{equation}
The NIR normalizations $A_{\rm NIR}^{\Delta E}$ are fixed by matching to the IP spectrum in \texttt{GB20}, and the IP normalizations in the two outer regions follow from the same condition in the 25--60\,keV band. The WD component is thus fully specified by $A_{\rm IP}^{\texttt{GB20}}$ and the three WD masses $M_{\rm WD}^{\texttt{GB20}}$, $M_{\rm WD}^{\texttt{L-80}}$, $M_{\rm WD}^{\texttt{L+80}}$, with all remaining quantities derived from Eq.~\eqref{eq:NIR_IP_consistency}. Both $M_{\rm WD}$ and $A_{\rm IP}$ are allowed to vary independently across the three sky regions.

\subsection{Fitting procedure and parameter space}
\label{subsec:fit}

The model prediction fitted to the 126 IBIS measurements consists of three components
\begin{equation}
    \boldsymbol{\Phi}_{\rm model}(\boldsymbol{\theta}) = \boldsymbol{\Phi}_{\rm DM}(\langle\sigma v\rangle \textrm{ or } \Gamma) + \boldsymbol{\Phi}_{\rm CR}(A_{\rm CR}) + \boldsymbol{\Phi}_{\rm IP}(A_{\rm IP}^{\texttt{GB20}}, M_{\rm WD}^{\rm regs})\;.
\end{equation}
For each science window, the predicted diffuse emission — from both the astrophysical backgrounds and the DM signal — is convolved with the IBIS/ISGRI collimator response (estimated in Ref.~\cite{Koechler:2026dci}) and averaged over the same set of observations used to construct the measured longitudinal profiles and energy spectra. Then, we define the likelihood as
\begin{equation}
    -2\log\mathcal{L}(\boldsymbol{\theta}) = \mathbf{R}(\boldsymbol{\theta})^\mathrm{T}\,\mathbf{C}^{-1}\,\mathbf{R}(\boldsymbol{\theta})\;,
    \label{eq:likelihood}
\end{equation}
where $\mathbf{R} = \boldsymbol{\Phi}_{\rm data} - \boldsymbol{\Phi}_{\rm model}$ and $\mathbf{C}$ is the covariance matrix estimated in Ref.~\cite{Koechler:2026dci}. The DM signal strength ($\langle\sigma v\rangle$ or $\Gamma$), $A_{\rm CR}$, and $A_{\rm IP}^{\texttt{GB20}}$ are treated as scale parameters and sampled in log-space, with log-uniform priors spanning $10^{-2}$--$10^{2}$ for $A_{\rm CR}$, and $10^{-10}$--$10^{2}$ for $A_{\rm IP}^{\texttt{GB20}}$, while the annihilation rate or lifetime are left unbounded. 
The WD masses are instead sampled linearly with a flat prior over the physically motivated range $[0.3, 1.4]\,M_\odot$, corresponding to the domain of validity of the IP spectral model~\cite{Suleimanov:2004aq}.

The astrophysical nuisance parameters can be partially degenerate with a DM signal, since background normalization and shape variations can absorb or mimic a putative contribution. To quantify the impact of this degeneracy, we consider three prior scenarios:

\textbf{Free priors case}: Log-uniform priors on $A_{\rm CR}$ and $A_{\rm IP}^{\texttt{GB20}}$; uniform priors on $M_{\rm WD}$.

$\mathbf{Case \,\, A_{\rm CR}\,\&\,M_{\rm WD}}$ \textbf{penalized} (benchmark): Gaussian penalties constrain $\log A_{\rm CR} \approx 0$ and $M_{\rm WD} \approx 0.6\,M_\odot$ with width $\sigma_{\rm pen} = 0.15$, motivated by the prior observational constraints on both quantities. $A_{\rm IP}^{\texttt{GB20}}$ retains a free prior given that it is an effective parameter not easily estimable.

$\mathbf{Case \,\, A_{\rm CR} = 1,\,M_{\rm WD}}$ \textbf{penalized}: $A_{\rm CR}$ is fixed to the model prediction, testing the impact of assuming a perfectly known CR normalization. This case relies more strongly on the model prediction, allowing us to assess the sensitivity of the derived limits to the assumption of a well-constrained background.

We carry out both Bayesian and frequentist analyses in order to better quantify how correlations between different parameters affect our results. In the Bayesian case, the posterior is sampled with \texttt{emcee}~\cite{Foreman-Mackey:2012any} and the $95\%$ upper limit is the 95th percentile of the marginalized DM posterior. 
In the frequentist case, we profile the likelihood over nuisance parameters using the differential-evolution minimizer~\cite{Storn:1997uea}, with the $95\%$ upper confidence limit defined by $\Delta\chi^2 = 2.71$. Our benchmark uses the penalized prior scenario without the estimated covariance matrix. Systematic uncertainties from the DM profile choice, propagation model parameters, prior scenario, covariance treatment, and statistical method are assessed separately and discussed in details in the Appendix~\ref{app:additional}.
We note that MCMC sampling is typically preferred over conventional frequentist confidence intervals, because it naturally accounts for parameter correlations and can capture non-Gaussian or asymmetric uncertainty distributions, which arises in nonlinear multi-parameter fits. The resulting credible intervals therefore provide a more complete characterization of the uncertainty in the fitted parameters. In this case, our results for the frequentist and Bayesian inference approaches yield roughly identical results.
We also perform a conservative analysis where we simply require that the DM signal does not exceed the data by more than two $\sigma$, applying the condition that $\chi^2_> < 4$ that has been used in previous X-ray constraints for DM~\cite{cirelli2023putting, Balaji:2025afr}.

\begin{figure}[t]
    \centering
    \includegraphics[width=\linewidth]{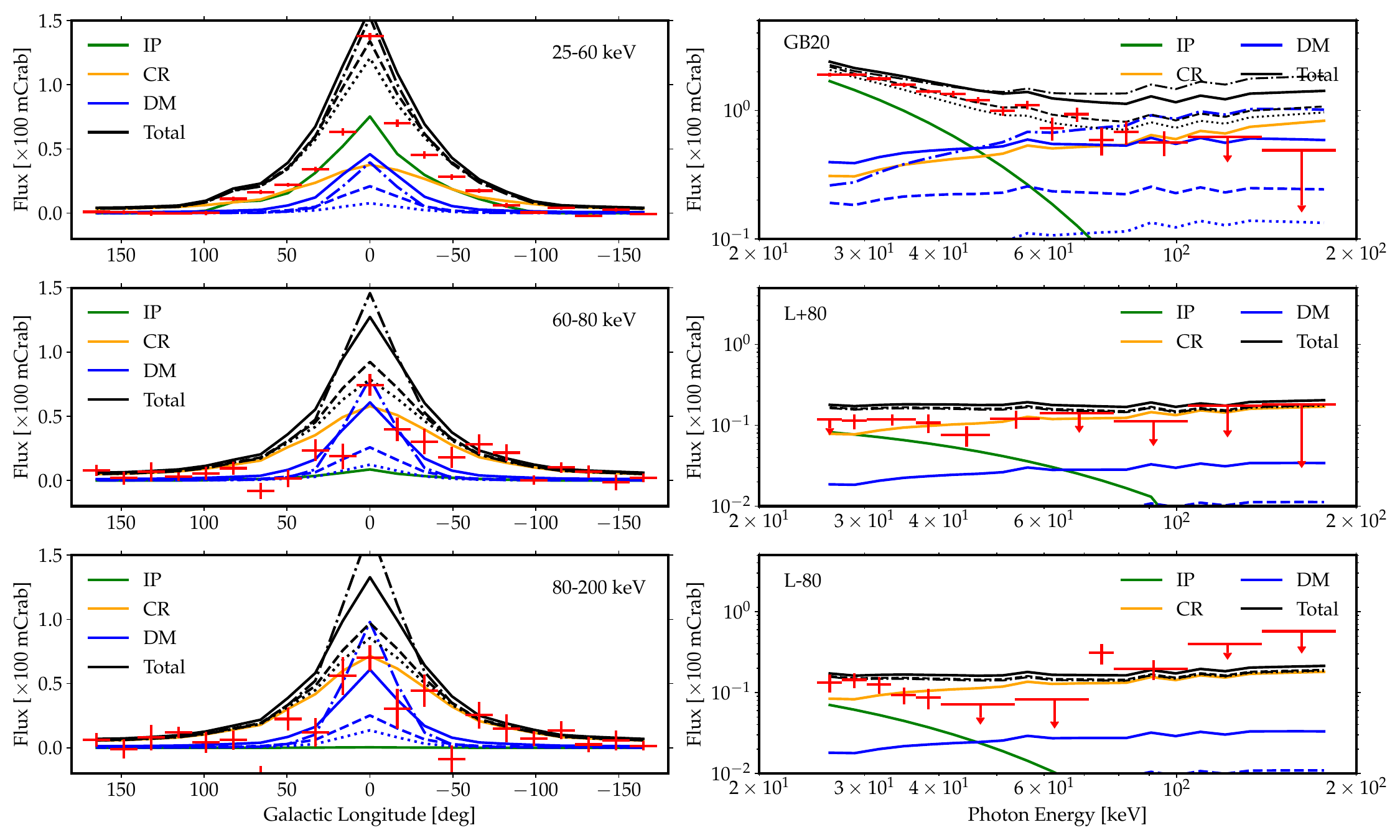}
    \caption{Morphology and spectra of the X-ray emissions, the red points being the measurements from IBIS/ISGRI. The DM-induced emissions from the annihilation into $e^+e^-$ (in blue) are from four cases: $m_{\rm DM} =1$ MeV and $\langle\sigma v\rangle=5\times10^{-29}$ cm$^3$/s (solid), $m_{\rm DM} =10$ MeV and $\langle\sigma v\rangle=5\times10^{-28}$ cm$^3$/s (dashed), $m_{\rm DM} =100$ MeV and $\langle\sigma v\rangle=1\times10^{-27}$ cm$^3$/s (dot-dashed), $m_{\rm DM} =1$ GeV and $\langle\sigma v\rangle=5\times10^{-27}$ cm$^3$/s (dotted). The predicted CR contribution is in orange, the IP contribution from WDs in green and the sum of all components is shown in black.}
    \label{fig:fluxmDMann}
\end{figure}

\section{Results: Predicted fluxes and morphologies}
\label{sec:results}

Figure~\ref{fig:fluxmDMann} shows the predicted spectra and longitudinal morphologies for the three sky regions defined in Section~\ref{sec:data}, compared to the IBIS measurements, for the case of DM annihilating into an $e^+e^-$ pair, and for $m_{\rm DM}=\{1, 10,100,1000\}$ MeV.
Several features are worth noting. For annihilation, the DM signal is more concentrated around the center of the Galaxy, with a morphology that reflects the squared DM density profile convolved with the CR propagation kernel. For decay, the signal morphology is more extended and less centrally concentrated (see Figure~\ref{fig:fluxmDMdec} in Appendix~\ref{app:additional}), reflecting the linear rather than quadratic dependence on the DM density. In both cases, the spectral shape of the DM-induced emission is more concentrated towards the center of the Galaxy than the CR component, with a harder spectrum than both the WD and CR components. 
The prompt FSR component contributes at lower energies and is more spatially concentrated toward the Galactic centre, closely following the DM density (or its square). However, this component remains subdominant with respect to the secondary emission at all the energies explored here. 


In a background-only fit, the astrophysical background model — CR electron IC emission (orange) and unresolved WD emission (green) — provides a reasonable description (Table I of Ref.~\cite{Koechler:2026dci}) of the data across all regions and energy bands, with the WD component dominating below $\sim60$\,keV and the CR component reproducing the high-energy tail above this threshold. The combined background-only fit is shown in Figure~\ref{fig:fluxnoDMBF} in Appendix~\ref{app:additional}. Interestingly, when the DM signal is included and $A_{\rm CR}$ is left free, the fit tends to redistribute the high-energy emission between the two components, preferring a subdominant CR contribution compensated by a non-negligible DM signal. This is driven by a preference in the fit for emission above $\sim60$~keV that is centrally concentrated and cannot be explained by the CR template. We show the best-fit values for the nuisance astrophysical parameters in Figure~\ref{fig:fitDM_MWDACR} in Appendix~\ref{App}.

\begin{figure}[t]
    \centering
    \includegraphics[width=0.49\linewidth]{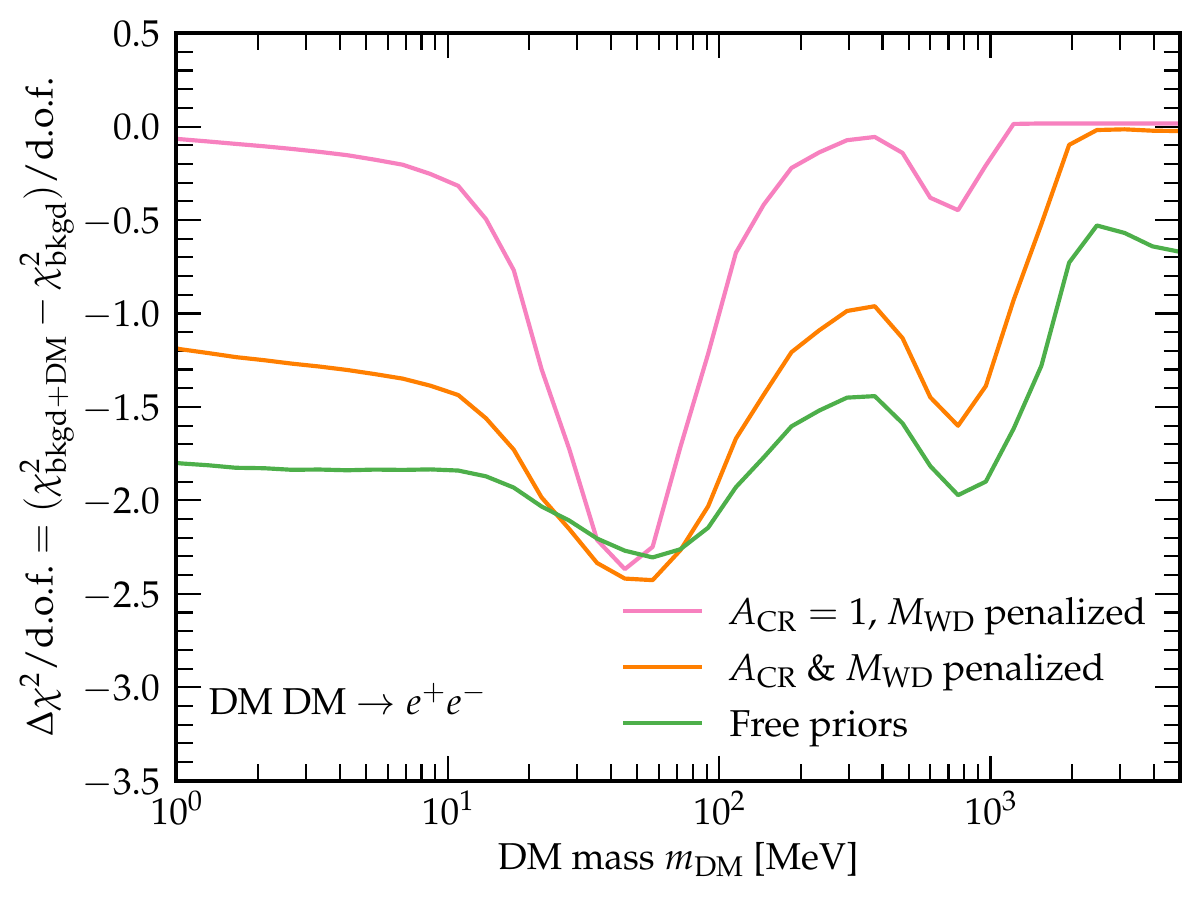}
    \includegraphics[width=0.49\linewidth]{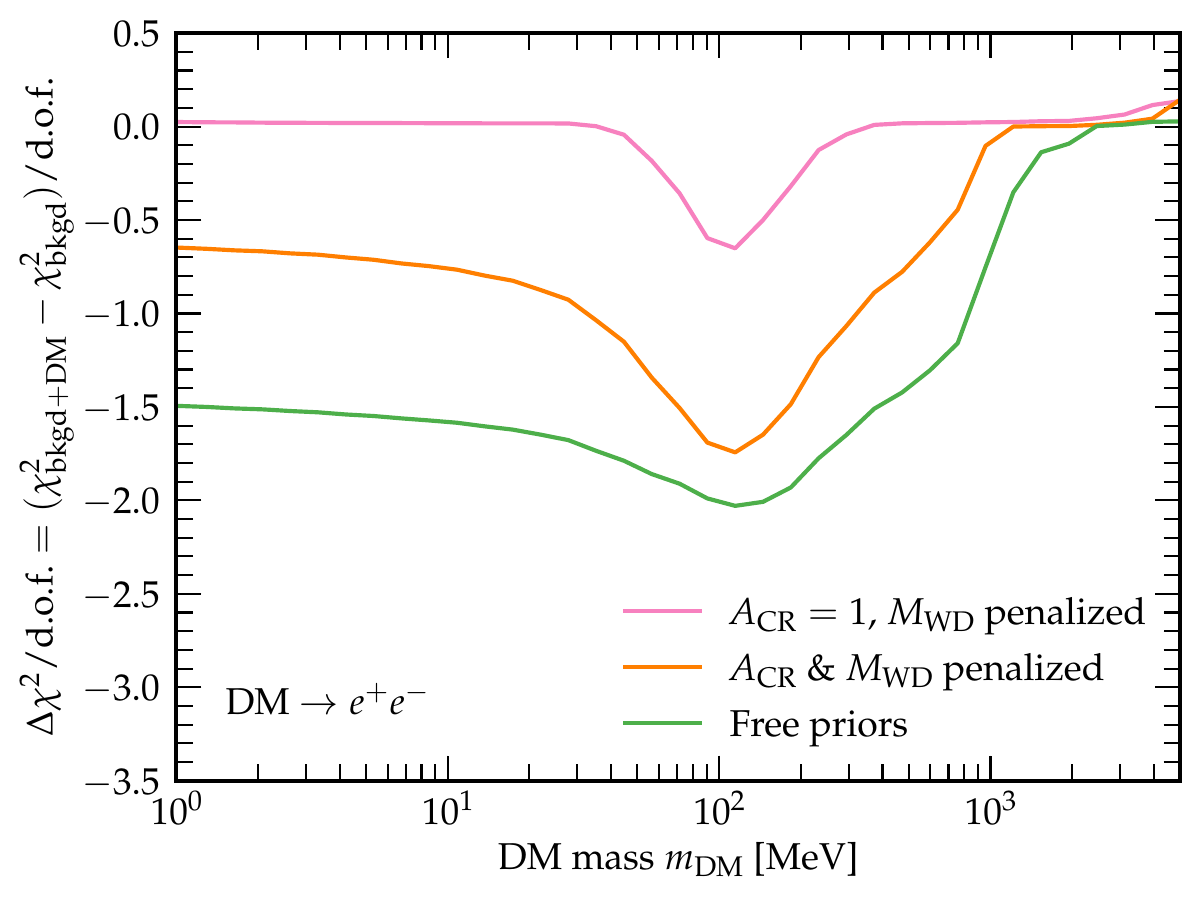}
    \caption{Difference in the goodness of fit between the DM+background and background-only hypotheses, per degree of freedom. The curves correspond to the three prior scenarios considered in the analysis, including correlations between observations. Negative values indicate that the PBH hypothesis provides a better fit. The left (right) panel shows the case of DM annihilating (decaying) in $e^+e^-$. The number of degrees of freedom here is 121.}
    \label{fig:chi2}
\end{figure}

To quantify the consistency of the data with the background-only hypothesis and assess whether an additional DM component is statistically preferred, we compare the best-fit $\chi^2$ values obtained under the two hypotheses for each DM mass and final state.
Figure~\ref{fig:chi2} shows the profile $\Delta\chi^2 = \chi^2_{\rm bkgd+DM} - \chi^2_{\rm bkgd}$ as a function of DM mass, for annihilation (left panel) and decay (right panel). A negative value indicates that including the DM component improves the fit over the background-only hypothesis. Remarkably, we find a consistent preference for an additional signal component across all prior scenarios and final states considered, with $\Delta\chi^2$ becoming significant around DM masses corresponding to electron injection energies of a few tens of MeV. This preference is present for both annihilation and decay, but is systematically more pronounced in the annihilation case — where the signal morphology scales as $\rho_{\rm DM}^2$ and is therefore more sharply concentrated toward the Galactic centre — suggesting that the data favour an additional emission component that is both spectrally soft and morphologically centrally peaked.

We stress that this result cannot be interpreted as evidence for a DM signal per se. The observed preference is equally consistent with an unmodelled astrophysical source injecting electrons at energies of a few tens of MeV in the inner Galaxy — a possibility that is independently motivated and of considerable interest in its own right. Indeed, the combination of the preferred injection energy scale and the morphological concentration toward the Galactic centre represents a potentially significant finding that warrants dedicated investigation; we defer a detailed study of this component and its possible astrophysical interpretations to an upcoming analysis. In this work, we conservatively treat the data as consistent with the background-only hypothesis and proceed to derive upper limits on the DM annihilation and decay rates.

\section{DM constraints}
\label{sec:constraints}

\begin{figure}[t]
    \centering
    \includegraphics[width=0.8\linewidth]{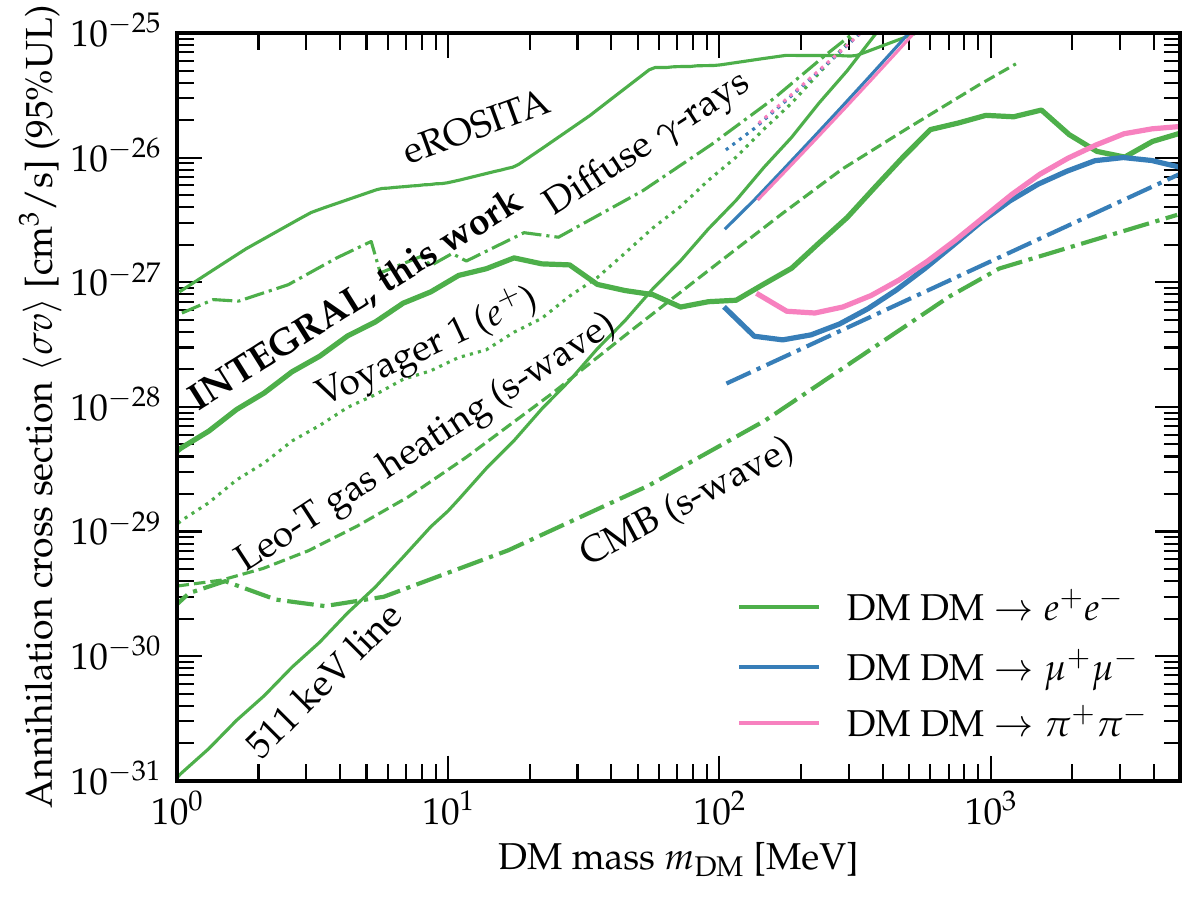}
    \caption{95\% upper limits on the bosonic DM annihilation cross section $\langle\sigma v\rangle$ as a function of DM mass $m_{\rm DM}$, for the three leptonic final states $e^+e^-$ (green), $\mu^+\mu^-$ (blue), and $\pi^+\pi^-$ (pink). The solid lines show the benchmark constraints derived in this work (INTEGRAL, this work), obtained with the $A_{\rm CR}\,\&\,M_{\rm WD}$ penalized prior and the frequentist approach without the full covariance matrix. Existing constraints from the CMB ($s$-wave)~\cite{Slatyer:2015jla,Lopez-Honorez:2013cua}, eROSITA~\cite{Balaji:2025afr}, Voyager~1~\cite{DelaTorreLuque:2023olp}, the Galactic 511\,keV line~\cite{DelaTorreLuque:2023cef}, Leo-T gas heating ($s$-wave)~\cite{Wadekar:2021qae}, and diffuse $\gamma$-rays~\cite{Essig:2013goa} are shown for comparison.}
    \label{fig:limitannDM}
\end{figure}

\begin{figure}[t]
    \centering
    \includegraphics[width=0.8\linewidth]{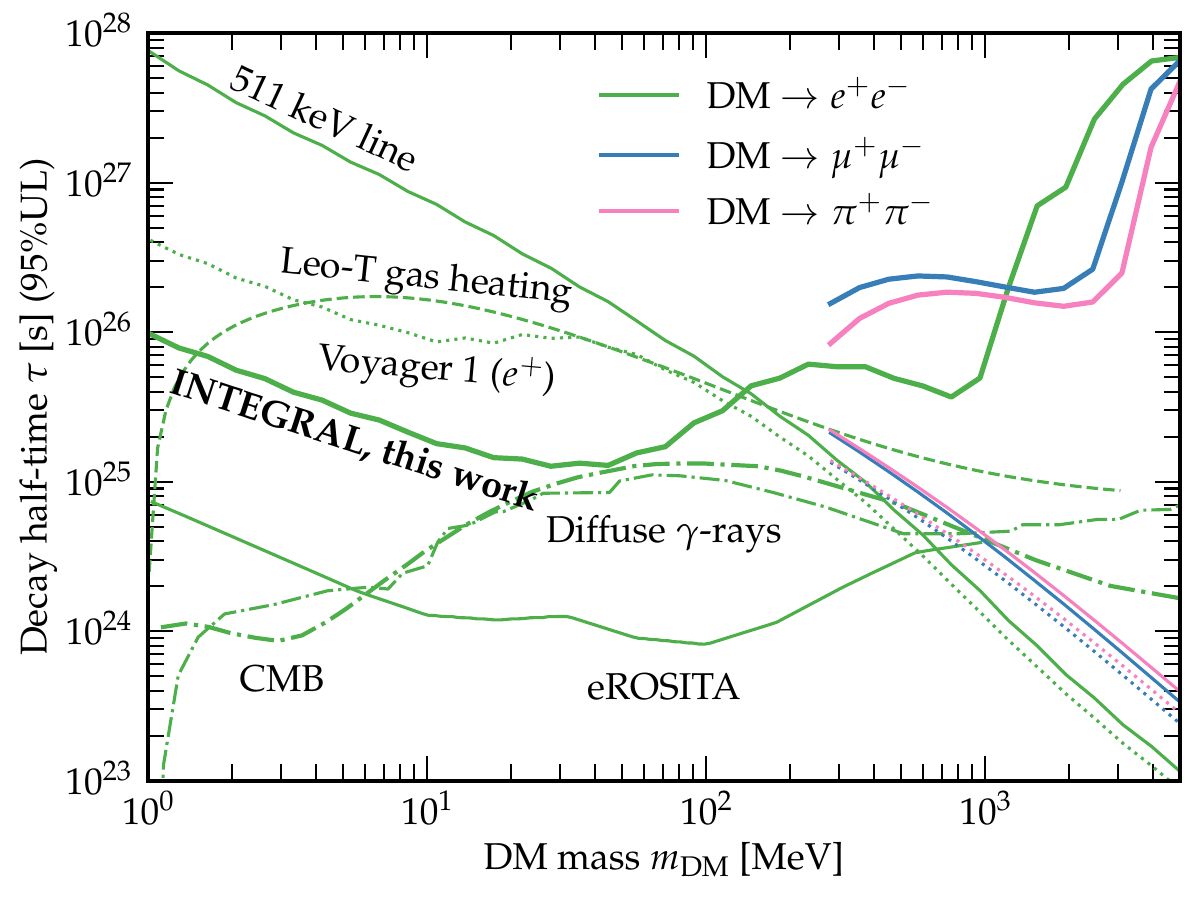}
    \caption{95\% upper limits on the DM decay half-life $\tau$ as a function of DM mass $m_{\rm DM}$, for the three leptonic final states $e^+e^-$ (green), $\mu^+\mu^-$ (blue), and $\pi^+\pi^-$ (pink). The solid lines show the benchmark constraints derived in this work (INTEGRAL, this work), obtained with the $A_{\rm CR}\,\&\,M_{\rm WD}$ penalized prior and the frequentist approach without the full covariance matrix. Existing constraints from the CMB~\cite{Liu:2016cnk}, eROSITA~\cite{Balaji:2025afr}, Voyager~1~\cite{DelaTorreLuque:2023olp}, the Galactic 511\,keV line~\cite{DelaTorreLuque:2023cef}, Leo-T gas heating~\cite{Wadekar:2021qae}, and diffuse $\gamma$-rays~\cite{Essig:2013goa} are shown for comparison.}
    \label{fig:limitdecDM}
\end{figure}

Our benchmark constraints on the bosonic DM annihilation cross section $\langle\sigma v\rangle$ and decay half-life $\tau$ are shown respectively in Figures~\ref{fig:limitannDM} and~\ref{fig:limitdecDM}, for the three leptonic final states $e^+e^-$, $\mu^+\mu^-$, and $\pi^+\pi^-$. These limits are derived using the $A_{\rm CR}\,\&\,M_{\rm WD}$ penalized prior and the frequentist profile-likelihood approach without including the full covariance matrix, as described in Section~\ref{sec:analysis}.

For DM annihilation (Figure~\ref{fig:limitannDM}), our constraints go up to $\langle\sigma v\rangle \lesssim 10^{-29}$\,cm$^3$\,s$^{-1}$ at DM masses of a few MeV for the $e^+e^-$ channel, improving by even more than two orders of magnitude over existing X-ray limits in the mass range where secondary IC emission dominates. Interestingly, the $\mu^+\mu^-$ and $\pi^+\pi^-$ channels, which are kinematically inaccessible at low masses, become more constraining than the $e^+e^-$ channel above their respective thresholds, where their harder prompt emission and boosted Michel spectra shift more of the signal into the observable energy range.
 Our limits are competitive with--and stronger at high masses than--those from CMB observations~\cite{Slatyer:2015jla} and Voyager~1~\cite{DelaTorreLuque:2023olp} measurements of local CR electrons, and significantly improve upon the eROSITA constraints~\cite{Balaji:2025afr} across the full mass range considered. The 511\,keV line constraint~\cite{DelaTorreLuque:2023cef} and Leo-T gas heating~\cite{Wadekar:2021qae} bounds provide complementary coverage at the lowest masses.

For DM decay (Figure~\ref{fig:limitdecDM}), our constraints reach half-lives of $10^{28}$ s $\gtrsim \tau \gtrsim 10^{25}$\,s across the sub-GeV mass range, again with the $e^+e^-$ channel providing the strongest bounds at low masses. The improvement over existing X-ray constraints is significant across the full mass range, with our limits surpassing eROSITA~\cite{Balaji:2025afr} by more than an order of magnitude. Constraints from the CMB~\cite{Liu:2016cnk}, Voyager~1~\cite{DelaTorreLuque:2023olp}, and diffuse $\gamma$-rays~\cite{Essig:2013goa} are shown for comparison; our limits are competitive with or stronger than these across most of the mass range, with the 511\,keV line~\cite{DelaTorreLuque:2023cef}, which provides the leading constraint at the lowest masses.

The robustness of these results with respect to the choice of prior scenario, the treatment of measurement correlations, the CR propagation model, and the DM density profile is assessed in detail in Appendix~\ref{App}. The dominant source of uncertainty is the CR propagation model, particularly the Alfvén velocity $v_A$ controlling diffusive reacceleration, which can shift the constraints by up to 3 orders of magnitude over the mass range considered. The DM profile, prior choice, and covariance treatment introduce comparatively smaller variations.

\subsection{Specific particle models}
\label{sec:particle_models}

The limits derived above are expressed in terms of the model-independent quantities $\langle\sigma v\rangle$ and $\Gamma = \tau^{-1}$. Their translation into constraints on microscopic portal couplings is discussed below for representative annihilating and decaying DM scenarios.

\subsubsection{Annihilating dark matter}
\label{subsec:annihilating_DM_models}

\paragraph{Dark photon.} A particularly well-motivated realization of sub-GeV DM is the dark-photon (or vector-portal) model. A new (dark) Abelian gauge symmetry $U(1)_D$ introduces a vector mediator $Z'_\mu$ that couples to the DM particle and communicates with the SM through kinetic mixing with hypercharge. The attractiveness of this scenario stems from its simplicity: the interaction is renormalizable, the small coupling to the SM is technically natural, and the mediator provides a well-defined portal connecting an otherwise secluded dark sector to ordinary matter. Moreover, the same interaction can generate both DM annihilation and DM--electron scattering through the mediator kinetic mixing, making the model particularly useful for connecting indirect and direct searches for sub-GeV DM. Such vector portals are therefore among the simplest benchmarks for light DM phenomenology~\cite{Cirelli:2026omb, Essig:2015cda}.

For definiteness, we consider DM as a Dirac fermion $\chi$ (with mass $m_\chi$) interacting with $Z'_\mu$ that couples through a kinetic mixing $\epsilon$ with the SM photon, with the following Lagrangian term:
\begin{equation}
    \label{eq:Ltot}
    \mathcal{L} \supset \bar\chi (i\gamma^\mu D_\mu-m_\chi)\chi-\frac{1}{4}F'_{\mu\nu}F'^{\mu\nu}+\frac{1}{2}m_{Z'}^2Z'_\mu Z'^\mu + \frac{\epsilon}{2}F'_{\mu\nu}B^{\mu\nu},
\end{equation}
where the covariant derivative is $D_\mu = \partial_\mu+ig_DZ'_\mu$, with $g_D$ is the coupling between DM and the $Z'$. $B_{\mu\nu}$ and $F'_{\mu\nu}$ are respectively the field-strength tensor of the hypercharge and the dark sector, while $m_{Z'}$ is the mass of the boson of the new $U(1)_D$, $Z'_\mu$.

The Lagrangian of Eq.~\ref{eq:Ltot} is not canonically written, and the mass of the SM bosons (which appear in the SM Lagrangian) and $m_{Z'}$ are not eigenmasses. After the electro-weak symmetry breaking and a proper rotation of the fields, the SM bosons and $Z'_\mu$ mix to give the photon, $Z$ boson and the dark photon $A'_\mu$. We obtain the following interaction Lagrangian term for the dark photon:
\begin{equation}
    \label{eq:Lcanon}
    \mathcal{L}_{\rm int} \approx - A'_\mu \left[g_D\bar\chi\gamma^\mu\chi-e\epsilon\cos\theta_W \sum_f \bar f\gamma^\mu f\right]\;,
\end{equation}
where $\theta_W$ is the Weinberg mixing angle.
From this we can work our way to the thermally averaged DM annihilation cross section $\langle\sigma v\rangle$ and the reference DM-electron scattering cross section $\bar\sigma_e$. 

The phenomenology depends qualitatively on whether the mediator is heavier or lighter than the DM particle. When the dark photon is heavier than the DM particle, the dominant annihilation process is $\chi\bar{\chi}\rightarrow A'^\ast\rightarrow f\bar f$. Away from the mediator resonance and in the limit $m_{A'}\gg m_\chi$, the annihilation cross section scales as \cite{DiMauro:2021qcf}
\begin{align}
\langle\sigma v\rangle &\simeq \frac{g_D^2e^2\epsilon^2m_\chi^2}{2\pi m_{A'}^4}\sum_f^{m_f<m_\chi} N_c^fQ_f^2\left(1+\frac{m_f^2}{2m_\chi^2}\right)\sqrt{1-\frac{m_f^2}{m_\chi^2}}+\mathcal{O}(v^2) \\
&\simeq \frac{8\pi\alpha y}{m_\chi^2} \sum_f^{m_f<m_\chi}\left(1+\frac{m_f^2}{2m_\chi^2}\right)\sqrt{1-\frac{m_f^2}{m_\chi^2}}+\mathcal{O}(v^2)\;,
\label{eq:vector_ann_scaling}
\end{align}
where $\alpha=e^2/(4\pi)$ is the fine-structure constant, $\alpha_D=g_D^2/(4\pi)$ is the dark fine-structure constant, $N_c^f$ is the color number of the fermion $f$ (1 for leptons and 3 for quarks), and $Q_f$ is its electric charge. Since we study the annihilation channels in $e^+e^-$, $\mu^+\mu^-$ and $\pi^+\pi^-$, the $Q_f=N_c^f=1$ for all of the studied final states. Finally, $y$ is defined as
\begin{equation}
y \equiv \alpha_D\epsilon^2 \left(\frac{m_\chi}{m_{A'}}\right)^4.
\label{eq:y_def}
\end{equation}

\begin{figure}[t]
    \centering
    \includegraphics[width=0.8\linewidth]{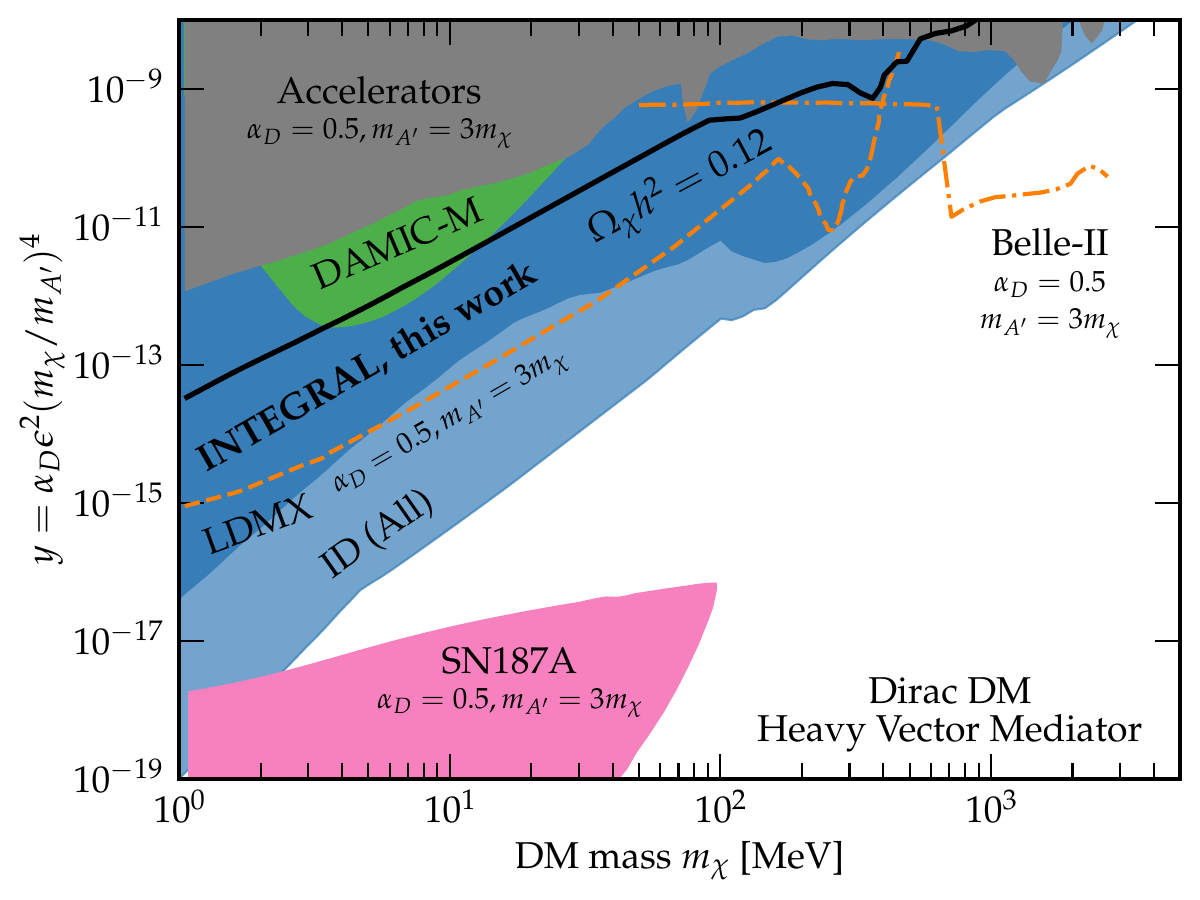}
    \caption{Limits on the dark photon mediator model, assuming $m_{A'}>m_\chi$. We show the limits derived in this work as the blue region, while the strongest indirect detection limit (consisting on the 511 keV line~\cite{DelaTorreLuque:2023cef} and the CMB~\cite{Slatyer:2015jla,Lopez-Honorez:2013cua} limits) is shown as the shaded blue region. The exclusion region from the absence of DM-induced electronic recoils in the DAMIC-M experiment~\cite{DAMIC-M:2025luv} is shown in green. The area of the parameter space excluded by accelerator searches is shown in grey~\cite{Krnjaic:2022ozp}, and the reach of Belle-II~\cite{Ferber:2022ewf} and the future LDMX experiment~\cite{LDMX:2018cma,LDMX:2025bog} is shown respectively in dot-dashed and dashed orange lines. Finally, the excluded region from the cooling time of SN1987A~\cite{Chang:2018rso} is shown in pink. The three latter limits are sensitive to the chosen values of $\alpha_D$ and $m_{A'}$, and are shown in the plot for $\alpha_D=0.5$ and $m_{A'}=3m_\chi$. The black line shows the thermal milestone~\cite{Krnjaic:2022ozp}.}
    \label{fig:limitannDM_DP}
\end{figure}

Since we separately computed the upper limits on $\langle\sigma v\rangle$ assuming all annihilation undergo into $e^+e^-$, $\mu^+\mu^-$ or $\pi^+\pi^-$, one needs to perform again the analysis by including the branching ratios into the final states, which will impact the spectral shape of the X-ray emissions. However, in this work we adopt a conservative approach, where the upper limit on the total annihilation cross section is the minimum one for the three channels
\begin{equation}
    \langle\sigma v\rangle^{\rm UL}\approx \min_f \langle\sigma v\rangle_{f\bar f}^{\rm UL}\;,
\end{equation}
where $\langle\sigma v\rangle_{f\bar f}^{\rm UL}$ is the derived UL assuming the annihilation undergoes solely into $f\bar f$. From this we can then translate into an estimated upper limit in the $(m_\chi-y)$ plane, as shown by the blue regions in Figure~\ref{fig:limitannDM_DP}.

In this same dark photon model, the reference DM-electron scattering cross section $\bar\sigma_e$ and the DM form factor $F_{\rm DM}$ can be written as~\cite{Essig:2015cda}:
\begin{equation}
    \bar\sigma_e = \frac{16\pi\mu_{\chi e}^2\alpha\alpha_D\epsilon^2}{(m_{A'}^2+\alpha^2 m_e^2)^2}\;,\quad F_{\rm DM}(q)=\frac{m_{A'}^2+\alpha^2m_e^2}{m_{A'}^2+q^2}\;,
\end{equation}
where $\mu_{\chi e}=m_\chi m_e/(m_\chi+m_e)$ is the reduced mass of the DM-electron system and $q$ is the momentum transfer. In the heavy mediator case, we can approximate these expressions:
\begin{equation}
    \bar\sigma_e \approx \frac{16\pi\mu_{\chi e}^2\alpha y}{m_\chi^4}\;,\quad F_{\rm DM}(q)\approx 1\;.
\end{equation}
We can therefore derive the exclusion in the $(m_\chi-y)$ plane from electronic recoil direct detection experiments, using the reported upper limit on $\bar\sigma_e$ (for $F_{\rm DM}(q)=1$) by the DAMIC-M experiment~\cite{DAMIC-M:2025luv}. This is shown in green in Figure~\ref{fig:limitannDM_DP}.

Strong constraints on the $(m_\chi-y)$ plane can be also set by accelerator experiments~\cite{BaBar:2017tiz,Andreev:2021fzd,Banerjee:2019pds,COHERENT:2021pvd} and the cooling time of the supernova SN1987A observed in the Large Magellanic Cloud~\cite{Chang:2018rso}. They are respectively shown in gray and pink in Figure~\ref{fig:limitannDM_DP}. Prospects of the fixed-target experiment LDMX~\cite{LDMX:2018cma,LDMX:2025bog}, and the sensitivity of the Belle-II experiment~\cite{Ferber:2022ewf} are shown with the orange lines~\cite{Krnjaic:2022ozp}. All of these constraints are sensitive to the adopted values of $\alpha_D$ and $m_{A'}$, and are shown for $\alpha_D=0.5$ and $m_{A'}=3m_\chi$. 

Figure~\ref{fig:limitannDM_DP} shows that indirect detection methods are essential in reaching (hence excluding) the thermal milestone in this dark matter model, \emph{i.e.}~the parameter space which sets the correct DM relic abundance.

The other kinematic limit of the dark photon model, is when the mediator is lighter DM particles ($m_{A'}<m_\chi$). In this case, the dominant DM annihilation channel becomes $\chi\bar\chi \to A'A'\to f\bar f f'\bar f'$. The upper limit on the associated annihilation cross section, $\langle\sigma v\rangle_{A'A'}$, cannot be directly inferred from the upper limits on $\langle\sigma v\rangle_{f\bar f}$ we derived. This needs a dedicated analysis that we leave for future work. We point the reader to Ref.~\cite{Cirelli:2026omb} for similar works.

\paragraph{Scalar mediator.} A second minimal possibility is a scalar mediator that mixes with the Higgs boson. For Dirac DM $\chi$ and a scalar mediator $\Phi$, the following Lagrangian terms are added to the SM one~\cite{Baek:2011aa}:
\begin{multline}
    \label{eq:Lsc}
    \mathcal{L}\supset \frac{1}{2}\partial_\mu \Phi\partial^\mu \Phi-\frac{1}{2}m_\Phi^2\Phi^2-\mu_\Phi^3\Phi-\frac{\mu'_\Phi}{3}\Phi^3-\frac{\lambda_\Phi}{4}\Phi^4+ \\+\bar\chi(i\gamma^\mu\partial_\mu-m_\chi)\chi-\lambda \Phi\bar\chi\chi-\mu_{H\Phi}\Phi H^\dagger H-\frac{\lambda_{H\Phi}}{2}\Phi^2H^\dagger H\;,
\end{multline}
where $H$ is the Higgs doublet. A proper rotation of the fields is a needed, and after electroweak symmetry breaking, the Higgs doublet mixes with $\Phi$ (and the $W$ and $Z$ bosons) to form a new scalar eigenstate $\phi$, and the SM Higgs boson. We obtain the following interaction Lagrandian term for the scalar $\phi$:
\begin{equation}
    \mathcal{L}_{\rm int}\approx -\phi\left[g_D \bar\chi\chi-\sin\theta\sum_f \frac{m_f}{v_h}\bar ff\right]\;,
\end{equation}
where $\theta$ is the mixing angle between the $H$ and $\Phi$ (which depends on the couplings in Eq.~\ref{eq:Lsc}), $g_D$ is the coupling between DM and $\phi$, and $v_h\simeq246$ GeV is the Higgs vacuum expectation value. Once again, we can write the expression of the DM annihilation cross section, which is the following when $m_\phi > m_\chi$ (where only $\chi\bar\chi\to\phi^\ast\to f\bar f$ is open):
\begin{align}
    \langle\sigma v\rangle &\simeq \frac{g_D^2m_\chi^2v^2\sin^2\theta}{8\pi v_h^2m_\phi^4}\sum_f^{m_f<m_\chi} m_f^2+\mathcal{O}(v^4)\;, \\
    &\simeq \frac{v^2y}{2v_h^2m_\chi^2}\sum_f^{m_f<m_\chi} m_f^2+\mathcal{O}(v^4)\;,
\end{align}
where $v$ is the relative velocity of DM particles in the Galaxy (for which we take $v\sim 10^{-3}$~\cite{Knapen:2017xzo}), and $y$ is defined as:
\begin{equation}
    y\equiv \alpha_D\sin^2\theta\left(\frac{m_\chi}{m_\phi}\right)^4\;.
\end{equation}

\begin{figure}[t]
    \centering
    \includegraphics[width=0.8\linewidth]{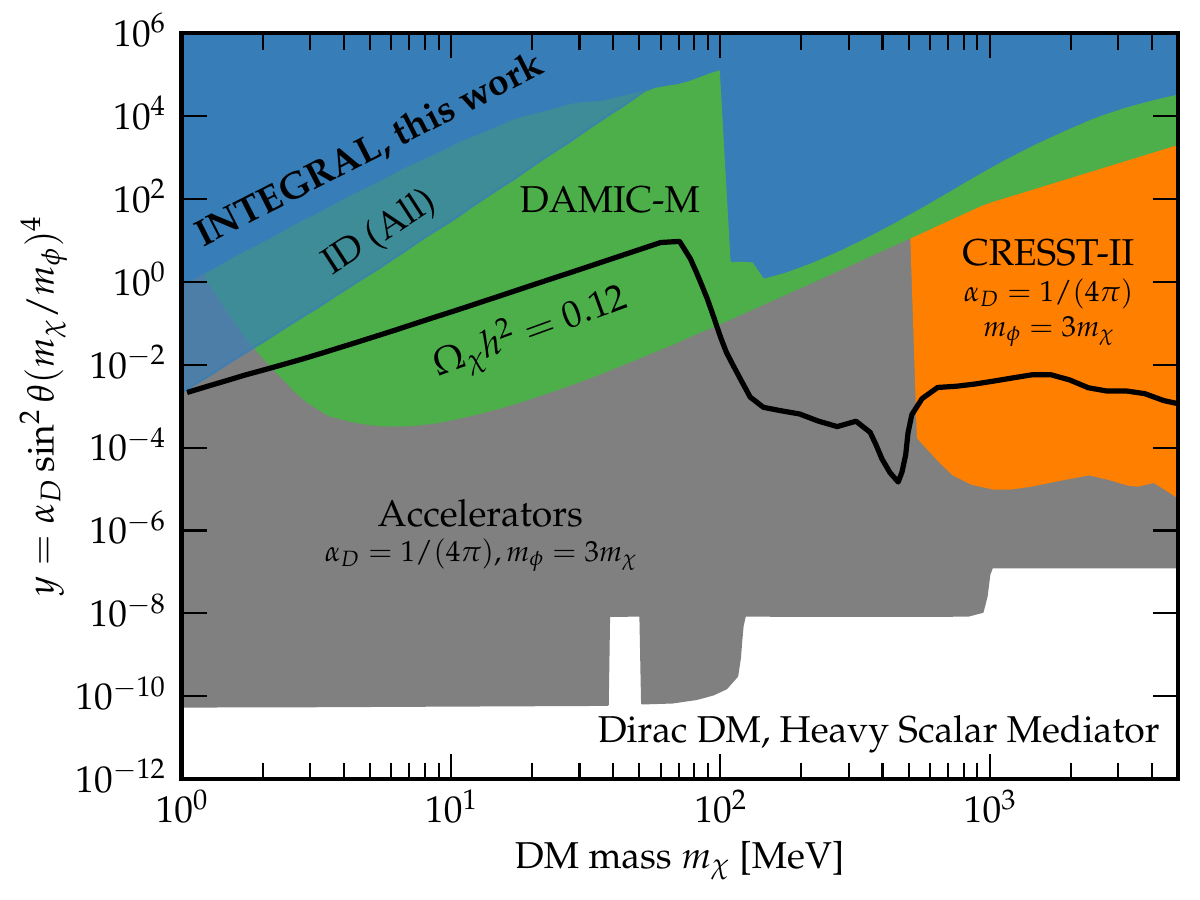}
    \caption{Limits on the scalar mediator model, assuming $m_\phi>m_\chi$. We show the limits derived in this work as the blue region, while the strongest indirect detection limit (consisting on the 511 keV line~\cite{DelaTorreLuque:2023cef} and our limits) is shown as the shaded blue region. The exclusion region from the absence of DM-induced electronic recoils in the DAMIC-M experiment~\cite{DAMIC-M:2025luv} is shown in green. The area of the parameter space excluded by accelerator searches is shown in grey~\cite{BaBar:2013npw,ATLAS:2015gvj}, and the exclusion from nuclear recoil direct detection CRESST-II~\cite{CRESST:2015txj} is shown in orange. The two latter limits are sensitive to the chosen values of $\alpha_D$ and $m_\phi$, and are shown in the plot for $\alpha_D=1/(4\pi)$ and $m_\phi=3m_\chi$. The black line shows the thermal milestone~\cite{Krnjaic:2015mbs}.}
    \label{fig:limitannDM_Sc}
\end{figure}

One can note that instead of being $s$-wave as in the case of the dark photon mediator, for the scalar mediator the annihilation cross section is $p$-wave. This already means that the limits from indirect detection on the model's parameter space are going to be sub-leading. We report these limits in the $(m_\phi-y)$ plane, shown as the blue regions in Figure~\ref{fig:limitannDM_Sc}.

We can also derive the reference DM-electron scattering cross section $\bar\sigma_e$ and DM form factor $F_{\rm DM}(q)$:
\begin{equation}
    \bar\sigma_e = \frac{4\mu_{\chi e}^2\alpha_Dm_e^2\sin^2\theta}{v_h^2(m_\phi^2+\alpha^2m_e^2)^2}\;, \quad F_{\rm DM}(q)=\frac{m_\phi^2+\alpha^2m_e^2}{m_\phi^2+q^2}\;,
\end{equation}
which can be simplified in the heavy mediator regime:
\begin{equation}
    \bar\sigma_e \approx \frac{4\mu_{\chi e}^2m_e^2y}{v_h^2m_\chi^4}\;, \quad F_{\rm DM}(q)\approx 1\;.
\end{equation}
We report the limits from DAMIC-M~\cite{DAMIC-M:2025luv} on the $(m_\phi-y)$ plane, as shown in the green region in Figure~\ref{fig:limitannDM_Sc}. Stronger constraints can be put from accelerator experiments~\cite{BaBar:2013npw,ATLAS:2015gvj} and the nuclear recoil direct detection experiment CRESST-II~\cite{CRESST:2015txj}. We report them respectively as grey and orange regions in Figure~\ref{fig:limitannDM_Sc}. These constraints are sensitive to the adopted values of $\alpha_D$ and $m_\phi$, and are shown for $\alpha_D=1/(4\pi)$ and $m_\phi=3m_\chi$. As already known, accelerators exclude the thermal milestone associated to the measured DM relic density, as they are insensitive to the velocity of DM particles. Electronic recoil experiments (such as DAMIC-M) also offers good complementarity to accelerators experiments in excluding this milestone.

\subsubsection{Decaying dark matter}
\label{subsec:decaying_DM_models}

Light bosonic DM is particularly well motivated in this context because a small portal coupling can naturally lead to macroscopic or cosmological lifetimes (therefore evading accelerator limits), while the decay products remain sufficiently energetic to produce observable electromagnetic signals~\cite{Nguyen:2025tkl, DelaTorreLuque:2025zjt, Balaji:2025alr, DelaTorreLuque:2024zsr}. We consider the dark photon and Higgs-mixed scalar as two minimal examples.

\begin{figure}[t]
    \centering
    \includegraphics[width=0.8\linewidth]{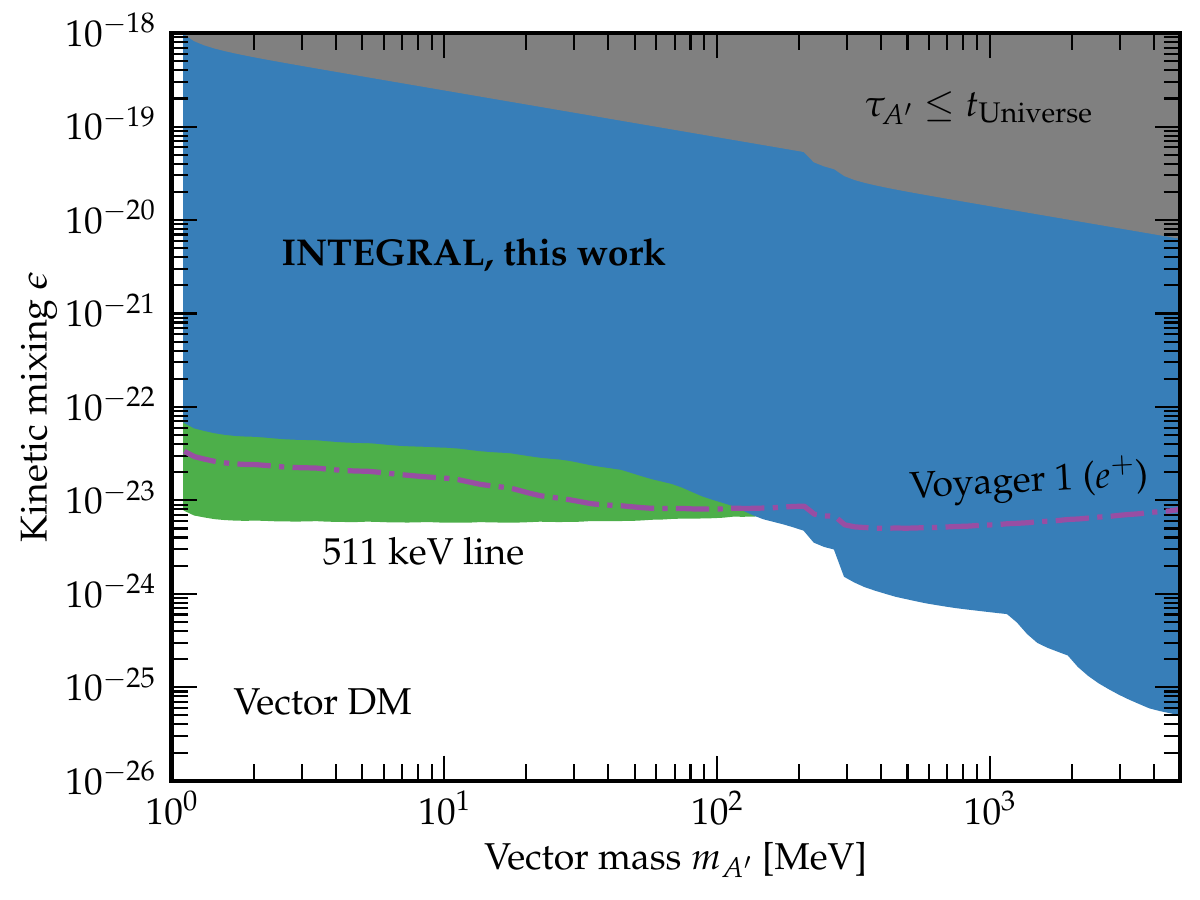}
    \caption{Limits on the decaying dark photon dark matter parameters space. We show the limits derived in this work as the blue region, while the 511 keV line limit is shown in green and the limit from Voyager-1 positron measurements is shown as the purple dot-dashed line. The parameter space in which the dark photon decays faster than the age of the Universe is shown in grey.}
    \label{fig:limitdecDM_DP}
\end{figure}

\paragraph{Dark photon.} The dark photon provides an especially simple example of a decaying dark-sector particle, since its interaction with the SM is uniquely determined by the kinetic mixing parameter $\epsilon$. The decay rate of the dark photon is written, assuming decays into $e^+e^-$, $\mu^+\mu^-$ and $\pi^+\pi^-$
\begin{equation}
    \Gamma=\tau^{-1}=\frac{\alpha\epsilon^2m_{A'}}{3}\sum_f^{2m_f<m_\chi}\left(1+\frac{2m_f^2}{m_{A'}^2}\right)\sqrt{1-\frac{4m_f^2}{m_{A'}^2}}\;.
\end{equation}
This gives a clear connection between the upper limits on $\tau$ we derived and an upper limit on the kinetic mixing $\epsilon$. Like the annihilation case, we assume conservatively that 
\begin{equation}
    \tau^{\rm UL}\approx \min_f \tau_{f\bar f}^{\rm UL}\;,
\end{equation}
instead of redoing the whole analysis with the correct branching fractions. The estimated exclusion of the $(m_{A'}-\epsilon)$ plane from our analysis is reported in blue in Figure~\ref{fig:limitdecDM_DP}.

In this Figure, we also report other indirect detection limits from the 511 keV line~\cite{DelaTorreLuque:2023cef} (green region) and Voyager-1 positron measurements~\cite{DelaTorreLuque:2023olp} (purple dot-dashed line). We display also in gray the parameter space in which the decay time of the dark photon is lower than the age of the Universe $t_{\rm Universe}\simeq 4.33\times10^{17}$ s.
Similarly to Figure~\ref{fig:limitdecDM}, it shows how stringent is the limit derived in our work compared to other limits from the literature, especially for DM masses above 200 MeV.

\begin{figure}[t]
    \centering
    \includegraphics[width=0.8\linewidth]{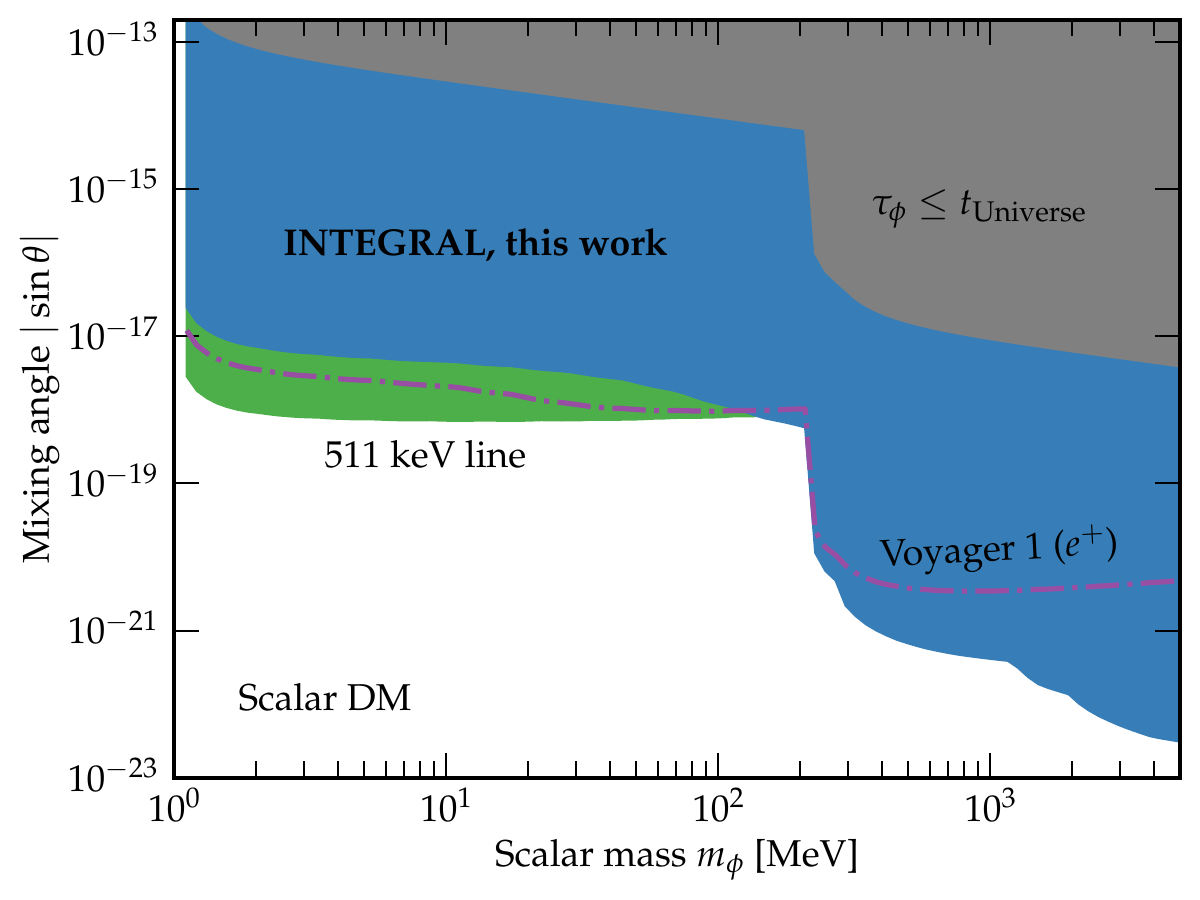}
    \caption{Limits on the decaying scalar dark matter parameters space. We show the limits derived in this work as the blue region, while the 511 keV line~\cite{DelaTorreLuque:2023cef} limit is shown in green and the limit from Voyager-1 positron measurements~\cite{DelaTorreLuque:2023olp} is shown as the purple dot-dashed line. The parameter space in which the dark photon decays faster than the age of the Universe is shown in grey.}
    \label{fig:limitdecDM_Sc}
\end{figure}

\paragraph{Scalar DM.} A scalar DM particle $\phi$ mixing with the Higgs boson acquires Yukawa-like couplings to SM fermions, with the coupling structure fixed by the SM Yukawa interactions and a small mixing angle naturally allowing a long lifetime. The decay rate of the scalar DM is written, assuming decays into $e^+e^-$, $\mu^+\mu^-$ and $\pi^+\pi^-$
\begin{equation}
\Gamma= \frac{m_\phi\sin^2\theta}{8\pi v_h^2}\sum_f^{2m_f<m_\chi} m_f^2\left(1-\frac{4m_f^2}{m_\phi^2}\right)^{3/2}\;.
\end{equation}
Our constraints on the $(m_\phi-|\sin\theta|)$ plane are shown in blue in Figure~\ref{fig:limitdecDM_Sc}. Like Figure~\ref{fig:limitdecDM_DP}, we report the limits from the 511 keV line~\cite{DelaTorreLuque:2023cef} (green region) and Voyager-1 positron measurements~\cite{DelaTorreLuque:2023olp} (purple dot-dashed line), and the DM stability criterion (gray).

\section{Discussion and conclusions}
\label{sec:conclusions}
 
In this work we have presented new constraints on sub-GeV dark matter annihilation and decay using twenty years of diffuse hard X-ray observations from the INTEGRAL/IBIS telescope. Our analysis extends the framework developed in our companion study of primordial black hole evaporation~\cite{Koechler:2026dci} to the case of particle DM interacting with the Standard Model, for a generic DM particle and also assuming couplings through vector and scalar portals, and represents the first joint spectral and morphological analysis of IBIS data in the context of light DM searches.
 
A central feature of our approach is the simultaneous exploitation of both the spectral shape and the spatial distribution of the diffuse hard X-ray emission. The 126 flux measurements — combining longitudinal profiles in three energy bands with energy spectra in three sky regions — encode complementary information: the spectrum constrains the energy of the injected electrons and therefore the DM mass and final state, while the morphology distinguishes signals concentrated toward the Galactic center (as expected from the NFW profile) from the more extended CR electron background. By fitting both simultaneously, we can be able to disentangle potential DM signals from the dominant astrophysical foregrounds far more effectively than spectral analyses alone, as demonstrated by the sensitivity of the $\Delta\chi^2$ curves to the spatial concentration of the signal.
 
We have constructed a physically motivated two-component background model combining IC emission from Galactic CR electrons — constrained by local and diffuse $\gamma$-ray data independently of this analysis — with thermal emission from unresolved accreting white dwarfs, modelled using the intermediate polar spectral template and the COBE/DIRBE near-infrared morphological template. The spectral and spatial descriptions of the WD component are tied together through a self-consistency condition, reducing the effective number of free parameters while preserving the physical coherence of the model. The normalizations of both components are left free in the fit, allowing the data to determine the relative contributions of each without imposing prior assumptions on their absolute levels.
 
Systematic uncertainties have been assessed across four axes: the choice of prior scenario for the nuisance parameters, the treatment of correlations between measurements via a cluster-bootstrap covariance matrix, the CR propagation model parametrized by the Alfvén velocity $v_A$, and the assumed DM density profile. The dominant uncertainty is CR propagation: varying $v_A$ from 0 to 40 km/s shifts the constraints by up to three orders of magnitude over the mass range considered, primarily because reacceleration modifies the energy and spatial distribution of the IC signal. The DM profile, prior scenario, and covariance treatment introduce comparatively smaller variations, confirming the robustness of our benchmark results against these aspects of the analysis.

 Intriguingly, across all prior scenarios and for both annihilation and decay final states considered, the data consistently prefer the inclusion of an additional diffuse emission component over the background-only hypothesis, with $\Delta\chi^2/{\rm d.o.f.}$ reaching values of order $-3$ to $-3.5$ around DM masses corresponding to electron injection energies of a few tens of MeV.
 The preference is systematically more pronounced in the annihilation case — where the signal morphology is more sharply peaked toward the Galactic center — than in the decay case. This strongly suggests that the additional component favoured by the data is not driven by spectral features alone, but points to the requirement for a centrally concentrated emission component that dominates the hard X-ray flux toward the Galactic center. Strikingly, the $\Delta\chi^2$ profiles for annihilation and decay both peak at the same characteristic electron injection energy, pinpointing a scale of a few tens of MeV as the one preferred by the data.
 We stress that this cannot be interpreted as evidence for a DM signal: the same morphological and spectral features could arise from an unmodelled astrophysical source injecting electrons at energies of a few tens of MeV in the Galactic center region, a possibility that is independently motivated by the long-standing excess in the 511\,keV line morphology~\cite{laTorreLuquePedro:2024est, DelaTorreLuque:2023cef}, the unexpectedly high ionization of hydrogen in the Central Molecular Zone~\cite{Oka:2005, Ravikularaman_2025, DelaTorreLuque:2024fcc}, and by the known difficulties in accounting for all low-energy CR electron sources in the inner Galaxy and the potential excess continuum $\gamma$-ray MeV emission~\cite{porter2008inverse, Orlando2018MNRAS.475.2724O, DelaTorreLuque:2024zsr, Balaji:2025alr}. The identification of this component — whether of astrophysical or exotic origin — and its possible connection to other Galactic centre anomalies represents an important open question that we defer to a dedicated upcoming analysis.
 
Therefore, we have conservatively interpreted this as a statistical fluctuation or an unmodelled astrophysical contribution, and have proceeded to derive upper limits. 
Our benchmark constraints reach $\langle\sigma v\rangle \lesssim 10^{-28}$\,cm$^3$\,s$^{-1}$ for DM annihilation into $e^+e^-$ at masses of a few MeV, improving by at least two orders of magnitude over existing X-ray limits from eROSITA~\cite{Balaji:2025afr} in the mass range where secondary IC emission dominates, and competitive with or stronger than the CMB~\cite{Slatyer:2015jla} and Voyager~1~\cite{DelaTorreLuque:2023olp} constraints over a broad mass range. For DM decay, our limits are particularly competitive at higher masses: these constraints reach $\tau \sim 10^{28}$\,s, surpassing eROSITA~\cite{Balaji:2025afr} by more than an order of magnitude in the whole mass range. Notably, in the decaying DM case our limits remain the leading X-ray constraints down to masses approaching $\sim1$\,GeV, where the 511 keV line begin to provide stronger bounds. This makes IBIS particularly valuable for probing the high-mass end of the sub-GeV window, where the IC signal from decaying DM falls in a regime that is challenging for both CMB and local CR measurements.
 
Additionally, we have translated our model-independent $\langle\sigma v\rangle$ and $\tau$ limits into constraints on the fundamental couplings of well-motivated DM models. For annihilating DM in the vector portal scenario ($\chi\bar\chi\to A'^\ast\to f\bar f$, for $m_{A'} > m_\chi$), our limits on $\langle\sigma v\rangle$ translate into constraints on the $y=\alpha_D\epsilon^2(m_\chi/m_{A'})^4$ parameter that can be directly compared with limits from direct DM--electron scattering experiments and collider searches. Similar translation was done for the scalar portal scenario ($\chi\bar\chi\to \phi^\ast\to f\bar f$, for $m_\phi > m_\chi$), where we show how our analysis can constrain the $y=\alpha_D\sin^2\theta(m_\chi/m_\phi)^4$ parameter. For decaying DM, we constrain the kinetic mixing parameter $\epsilon$ of a dark photon and the scalar-Higgs mixing angle $|\sin\theta|$ of a Higgs-portal scalar, providing limits that are complementary to those from accelerator-based experiments searching for long-lived dark sector particles. These comparisons demonstrate that indirect X-ray constraints from IBIS probe a region of coupling-mass parameter space that is not yet fully covered by direct detection, collider, or CMB measurements. 
 
Several directions remain open for future work. The hint for an additional centrally concentrated emission component at electron injection energies of a few tens of MeV 
warrants dedicated investigation, both in terms of its possible astrophysical origin and its potential connection to other anomalies associated to the low energies and the Galactic center. On the analysis side, extending the morphological information beyond longitudinal profiles to two-dimensional sky maps would further improve the signal-to-background discrimination, as would including additional energy bands where forthcoming missions, like COSI~\cite{tomsick2019comptonspectrometerimager} will provide improved sensitivity. Finally, extending the analysis from the dark photon portal model to models that gauge the SM accidental symmetries of $B-L$ and $L_i-L_j$, which couple directly to well-defined SM currents without relying on kinetic mixing, would broaden the particle-physics scope of these constraints.

\acknowledgments

We thank Roman Krivonos for helpful discussions about the data reduction and the astrophysical models. This research is based on observations with INTEGRAL, an ESA project with instruments and science data centre funded by ESA member states (especially the PI countries: Denmark, France, Germany, Italy, Switzerland, Spain), Czech Republic and Poland, and with the participation of Russia and the USA. P.D.L.~has been supported by the Juan de la Cierva JDC2022-048916-I grant, funded by MCIU/AEI/10.13039/501100011033 European Union "NextGenerationEU"/PRTR, and is currently supported by Ramón y Cajal RYC2024-048445-I grant, which is funded by MCIU/AEI/10.13039/501100011033 and FSE+. The work of P.D.L.~is also supported by the grants PID2021-125331NB-I00 and CEX2020-001007-S, both funded by MCIN/AEI/10.13039/501100011033 and by “ERDF A way of making Europe”. P.D.L.~also acknowledges the MultiDark Network, ref. RED2022-134411-T. J.K.~acknowledges support from the research grant {\sl TAsP (Theoretical Astroparticle Physics)} funded by Istituto Nazionale di Fisica Nucleare (INFN), and from the Italian Ministry of University and Research (MUR), PRIN 2022 ``EXSKALIBUR – Euclid-Cross-SKA: Likelihood Inference Building for Universe’s Research'', Grant No. 20222BBYB9, CUP I53D23000610 0006, and from the European Union -- Next Generation EU. J.K.~ also acknowledges support from the Italian Space Agency through the ASI INFN agreement n. 2018-28-HH.0: “Partecipazione italiana al GAPS - General AntiParticle Spectrometer”.

\bibliographystyle{JHEP.bst}
\bibliography{paper.bib}

\providecommand{\href}[2]{#2}\begingroup\raggedright\begin{thebibliography}{10}

\bibitem{Cirelli:2020bpc}
M.~Cirelli et~al., \emph{{Integral X-ray constraints on sub-GeV Dark Matter}}, \href{https://doi.org/10.1103/PhysRevD.103.063022}{\emph{Phys. Rev. D} {\bfseries 103} (2021) 063022} [\href{https://arxiv.org/abs/2007.11493}{{\ttfamily 2007.11493}}].

\bibitem{Cirelli:2023tnx}
M.~Cirelli et~al., \emph{{Putting all the X in one basket: Updated X-ray constraints on sub-GeV Dark Matter}}, \href{https://doi.org/10.1088/1475-7516/2023/07/026}{\emph{JCAP} {\bfseries 07} (2023) 026} [\href{https://arxiv.org/abs/2303.08854}{{\ttfamily 2303.08854}}].

\bibitem{DelaTorreLuque:2023olp}
P.~De~la Torre~Luque, S.~Balaji and J.~Koechler, \emph{{Importance of Cosmic-Ray Propagation on Sub-GeV Dark Matter Constraints}}, \href{https://doi.org/10.3847/1538-4357/ad41e0}{\emph{Astrophys. J.} {\bfseries 968} (2024) 46} [\href{https://arxiv.org/abs/2311.04979}{{\ttfamily 2311.04979}}].

\bibitem{Balaji:2025afr}
S.~Balaji, D.~Cleaver, P.~De~la Torre~Luque and M.~Michailidis, \emph{{Dark Matter in X-rays: Revised XMM-Newton Limits and New Constraints from eROSITA}}, \href{https://doi.org/10.1088/1475-7516/2025/11/053}{\emph{JCAP} {\bfseries 2025} (2025) 053} [\href{https://arxiv.org/abs/2506.02310}{{\ttfamily 2506.02310}}].

\bibitem{Krivonos:2006px}
R.~Krivonos, M.~Revnivtsev, E.~Churazov, S.~Sazonov, S.~Grebenev and R.~Sunyaev, \emph{{Hard X-ray emission from the Galactic ridge}}, \href{https://doi.org/10.1051/0004-6361:20065626}{\emph{A\&A} {\bfseries 463} (2007) 957} [\href{https://arxiv.org/abs/astro-ph/0605420}{{\ttfamily astro-ph/0605420}}].

\bibitem{Ubertini:2003ih}
P.~Ubertini et~al., \emph{{IBIS: The Imager on-board INTEGRAL}}, \href{https://doi.org/10.1051/0004-6361:20031224}{\emph{A\&A} {\bfseries 411} (2003) L131}.

\bibitem{Krivonos:2020qvl}
R.~Krivonos et~al., \emph{{$NuSTAR$ measurement of the cosmic X-ray background in the 3\textendash{}20 keV energy band}}, \href{https://doi.org/10.1093/mnras/stab209}{\emph{Mon. Not. Roy. Astron. Soc.} {\bfseries 502} (2021) 3966} [\href{https://arxiv.org/abs/2011.11469}{{\ttfamily 2011.11469}}].

\bibitem{Koechler:2026dci}
J.~Koechler and P.~De~la Torre~Luque, \emph{{Constraining dark matter using 20-year INTEGRAL/IBIS observations I: Primordial black holes}},  \href{https://arxiv.org/abs/2609.12044}{{\ttfamily 2609.12044}}.

\bibitem{Krivonos:2024bih}
R.~Krivonos, E.~Shtykovskaya and S.~Sazonov, \emph{{The properties of the Galactic hard X-ray and soft {\ensuremath{\gamma}}-ray background based on 20 years of INTEGRAL/IBIS observations}}, \href{https://doi.org/10.1016/j.jheap.2024.11.014}{\emph{JHEAp} {\bfseries 45} (2025) 96} [\href{https://arxiv.org/abs/2409.20058}{{\ttfamily 2409.20058}}].

\bibitem{LEDOIT2004365}
O.~Ledoit and M.~Wolf, \emph{A well-conditioned estimator for large-dimensional covariance matrices}, \href{https://doi.org/https://doi.org/10.1016/S0047-259X(03)00096-4}{\emph{Journal of Multivariate Analysis} {\bfseries 88} (2004) 365}.

\bibitem{Navarro:1995iw}
J.F.~Navarro, C.S.~Frenk and S.D.~White, \emph{{The Structure of cold dark matter halos}}, \href{https://doi.org/10.1086/177173}{\emph{Astrophys. J.} {\bfseries 462} (1996) 563} [\href{https://arxiv.org/abs/astro-ph/9508025}{{\ttfamily astro-ph/9508025}}].

\bibitem{Michel:1950}
L.~Michel, \emph{{Interaction between four half-spin particles and the decay of the $\mu$-meson}}, {\emph{Proc. Phys. Soc. A} {\bfseries 63} (1950) 514}.

\bibitem{Donato:2003xg}
F.~Donato, N.~Fornengo, D.~Maurin and P.~Salati, \emph{{Antiprotons in cosmic rays from neutralino annihilation}}, \href{https://doi.org/10.1103/PhysRevD.69.063501}{\emph{Phys. Rev. D} {\bfseries 69} (2004) 063501} [\href{https://arxiv.org/abs/astro-ph/0306207}{{\ttfamily astro-ph/0306207}}].

\bibitem{DRAGON2-1}
C.~Evoli et~al., \emph{{Cosmic-ray propagation with $\small{DRAGON2}$: I. numerical solver and astrophysical ingredients}}, \href{https://doi.org/10.1088/1475-7516/2017/02/015}{\emph{JCAP} {\bfseries 02} (2017) 015} [\href{https://arxiv.org/abs/1607.07886}{{\ttfamily 1607.07886}}].

\bibitem{DRAGON2-2}
C.~Evoli et~al., \emph{{Cosmic-ray propagation with DRAGON2: II. Nuclear interactions with the interstellar gas}}, \href{https://doi.org/10.1088/1475-7516/2018/07/006}{\emph{JCAP} {\bfseries 07} (2018) 006} [\href{https://arxiv.org/abs/1711.09616}{{\ttfamily 1711.09616}}].

\bibitem{Dundovic:2021ryb}
A.~Dundovic, C.~Evoli, D.~Gaggero and D.~Grasso, \emph{{Simulating the Galactic multi-messenger emissions with HERMES}}, \href{https://doi.org/10.1051/0004-6361/202140801}{\emph{Astron. Astrophys.} {\bfseries 653} (2021) A18} [\href{https://arxiv.org/abs/2105.13165}{{\ttfamily 2105.13165}}].

\bibitem{Churazov:2006bk}
E.~Churazov et~al., \emph{{INTEGRAL observations of the cosmic X-ray background in the 5-100 keV range via occultation by the Earth}}, \href{https://doi.org/10.1051/0004-6361:20066230}{\emph{A\&A} {\bfseries 467} (2007) 529} [\href{https://arxiv.org/abs/astro-ph/0608250}{{\ttfamily astro-ph/0608250}}].

\bibitem{delaTorreLuque:2022vhm}
P.~de~la Torre~Luque, M.N.~Mazziotta, A.~Ferrari, F.~Loparco, P.~Sala and D.~Serini, \emph{{FLUKA cross sections for cosmic-ray interactions with the DRAGON2 code}}, \href{https://doi.org/10.1088/1475-7516/2022/07/008}{\emph{JCAP} {\bfseries 07} (2022) 008} [\href{https://arxiv.org/abs/2202.03559}{{\ttfamily 2202.03559}}].

\bibitem{DelaTorreLuque:2023zyd}
P.~De~la Torre~Luque, F.~Loparco and M.N.~Mazziotta, \emph{{The FLUKA cross sections for cosmic-ray leptons and uncertainties on current positron predictions}}, \href{https://doi.org/10.1088/1475-7516/2023/10/011}{\emph{JCAP} {\bfseries 10} (2023) 011} [\href{https://arxiv.org/abs/2305.02958}{{\ttfamily 2305.02958}}].

\bibitem{Casandjian:2015hja}
J.-M.~Casandjian, \emph{{Local HI emissivity measured with Fermi-LAT and implications for cosmic-ray spectra}}, \href{https://doi.org/10.1088/0004-637X/806/2/240}{\emph{APJ} {\bfseries 806} (2015) 240} [\href{https://arxiv.org/abs/1506.00047}{{\ttfamily 1506.00047}}].

\bibitem{2016PhRvD..94f3009V}
S.~{Vernetto} and P.~{Lipari}, \emph{{Absorption of very high energy gamma rays in the Milky Way}}, \href{https://doi.org/10.1103/PhysRevD.94.063009}{\emph{PRD} {\bfseries 94} (2016) 063009} [\href{https://arxiv.org/abs/1608.01587}{{\ttfamily 1608.01587}}].

\bibitem{laTorreLuquePedro:2024est}
D.~la~Torre Luque;~Pedro, S.~Balaji, M.~Fairbairn, F.~Sala and J.~Silk, \emph{{511 keV galactic photons from a dark matter spike}}, \href{https://doi.org/10.1088/1475-7516/2025/09/034}{\emph{JCAP} {\bfseries 09} (2025) 034} [\href{https://arxiv.org/abs/2410.16379}{{\ttfamily 2410.16379}}].

\bibitem{DelaTorreLuque:2024zsr}
P.~De~la Torre~Luque, S.~Balaji, P.~Carenza and L.~Mastrototaro, \emph{{{\ensuremath{\gamma}} rays from in-flight positron annihilation as a probe of new physics}}, \href{https://doi.org/10.1103/PhysRevD.111.L061303}{\emph{Phys. Rev. D} {\bfseries 111} (2025) L061303} [\href{https://arxiv.org/abs/2405.08482}{{\ttfamily 2405.08482}}].

\bibitem{Suleimanov:2004aq}
V.~Suleimanov, M.~Revnivtsev and H.~Ritter, \emph{{Rxte broadband x-ray spectra of intermediate polars and white dwarf mass estimates}}, \href{https://doi.org/10.1051/0004-6361:20041283}{\emph{Astron. Astrophys.} {\bfseries 435} (2005) 191} [\href{https://arxiv.org/abs/astro-ph/0405236}{{\ttfamily astro-ph/0405236}}].

\bibitem{Turler:2010pm}
M.~Turler, M.~Chernyakova, T.J.L.~Courvoisier, P.~Lubinski, A.~Neronov, N.~Produit et~al., \emph{{INTEGRAL hard X-ray spectra of the cosmic X-ray background and Galactic ridge emission}}, \href{https://doi.org/10.1051/0004-6361/200913072}{\emph{Astron. Astrophys.} {\bfseries 512} (2010) A49} [\href{https://arxiv.org/abs/1001.2110}{{\ttfamily 1001.2110}}].

\bibitem{Yuasa:2012qe}
T.~Yuasa, K.~Makishima and K.~Nakazawa, \emph{{Broad-band spectral analysis of the Galactic Ridge X-ray Emission}}, \href{https://doi.org/10.1088/0004-637X/753/2/129}{\emph{Astrophys. J.} {\bfseries 753} (2012) 129} [\href{https://arxiv.org/abs/1205.1574}{{\ttfamily 1205.1574}}].

\bibitem{Foreman-Mackey:2012any}
D.~Foreman-Mackey, D.W.~Hogg, D.~Lang and J.~Goodman, \emph{{emcee: The MCMC Hammer}}, \href{https://doi.org/10.1086/670067}{\emph{Publ. Astron. Soc. Pac.} {\bfseries 125} (2013) 306} [\href{https://arxiv.org/abs/1202.3665}{{\ttfamily 1202.3665}}].

\bibitem{Storn:1997uea}
R.~Storn and K.~Price, \emph{{Differential Evolution {\textendash} A Simple and Efficient Heuristic for global Optimization over Continuous Spaces}}, \href{https://doi.org/10.1023/A:1008202821328}{\emph{J. Global Optim.} {\bfseries 11} (1997) 341}.

\bibitem{cirelli2023putting}
M.~Cirelli et~al., \emph{Putting all the x in one basket: Updated x-ray constraints on sub-gev dark matter},  2023.

\bibitem{Slatyer:2015jla}
T.R.~Slatyer, \emph{{Indirect dark matter signatures in the cosmic dark ages. I. Generalizing the bound on s-wave dark matter annihilation from Planck results}}, \href{https://doi.org/10.1103/PhysRevD.93.023527}{\emph{Phys. Rev. D} {\bfseries 93} (2016) 023527} [\href{https://arxiv.org/abs/1506.03811}{{\ttfamily 1506.03811}}].

\bibitem{Lopez-Honorez:2013cua}
L.~Lopez-Honorez et~al., \emph{{Constraints on dark matter annihilation from CMB observationsbefore Planck}}, \href{https://doi.org/10.1088/1475-7516/2013/07/046}{\emph{JCAP} {\bfseries 07} (2013) 046} [\href{https://arxiv.org/abs/1303.5094}{{\ttfamily 1303.5094}}].

\bibitem{DelaTorreLuque:2023cef}
P.~De~la Torre~Luque, S.~Balaji and J.~Silk, \emph{{New 511 keV Line Data Provide Strongest sub-GeV Dark Matter Constraints}}, \href{https://doi.org/10.3847/2041-8213/ad72f4}{\emph{Astrophys. J. Lett.} {\bfseries 973} (2024) L6} [\href{https://arxiv.org/abs/2312.04907}{{\ttfamily 2312.04907}}].

\bibitem{Wadekar:2021qae}
D.~Wadekar and Z.~Wang, \emph{{Strong constraints on decay and annihilation of dark matter from heating of gas-rich dwarf galaxies}}, \href{https://doi.org/10.1103/PhysRevD.106.075007}{\emph{Phys. Rev. D} {\bfseries 106} (2022) 075007} [\href{https://arxiv.org/abs/2111.08025}{{\ttfamily 2111.08025}}].

\bibitem{Essig:2013goa}
R.~Essig, E.~Kuflik, S.D.~McDermott, T.~Volansky and K.M.~Zurek, \emph{{Constraining Light Dark Matter with Diffuse X-Ray and Gamma-Ray Observations}}, \href{https://doi.org/10.1007/JHEP11(2013)193}{\emph{JHEP} {\bfseries 11} (2013) 193} [\href{https://arxiv.org/abs/1309.4091}{{\ttfamily 1309.4091}}].

\bibitem{Liu:2016cnk}
H.~Liu, T.R.~Slatyer and J.~Zavala, \emph{{Contributions to cosmic reionization from dark matter annihilation and decay}}, \href{https://doi.org/10.1103/PhysRevD.94.063507}{\emph{Phys. Rev. D} {\bfseries 94} (2016) 063507} [\href{https://arxiv.org/abs/1604.02457}{{\ttfamily 1604.02457}}].

\bibitem{Cirelli:2026omb}
M.~Cirelli, A.~Kar, A.~Letessier-Selvon and H.~Lumengo-Kidimbu, \emph{{Direct and Indirect searches for DM-electron interactions in sub-GeV DM models}},  \href{https://arxiv.org/abs/2608.26467}{{\ttfamily 2608.26467}}.

\bibitem{Essig:2015cda}
R.~Essig, M.~Fernandez-Serra, J.~Mardon, A.~Soto, T.~Volansky and T.-T.~Yu, \emph{{Direct Detection of sub-GeV Dark Matter with Semiconductor Targets}}, \href{https://doi.org/10.1007/JHEP05(2016)046}{\emph{JHEP} {\bfseries 05} (2016) 046} [\href{https://arxiv.org/abs/1509.01598}{{\ttfamily 1509.01598}}].

\bibitem{DiMauro:2021qcf}
M.~Di~Mauro and M.W.~Winkler, \emph{{Multimessenger constraints on the dark matter interpretation of the Fermi-LAT Galactic center excess}}, \href{https://doi.org/10.1103/PhysRevD.103.123005}{\emph{Phys. Rev. D} {\bfseries 103} (2021) 123005} [\href{https://arxiv.org/abs/2101.11027}{{\ttfamily 2101.11027}}].

\bibitem{DAMIC-M:2025luv}
{\scshape DAMIC-M} collaboration, \emph{{Probing Benchmark Models of Hidden-Sector Dark Matter with DAMIC-M}}, \href{https://doi.org/10.1103/2tcc-bqck}{\emph{Phys. Rev. Lett.} {\bfseries 135} (2025) 071002} [\href{https://arxiv.org/abs/2503.14617}{{\ttfamily 2503.14617}}].

\bibitem{Krnjaic:2022ozp}
G.~Krnjaic et~al., \emph{{A Snowmass Whitepaper: Dark Matter Production at Intensity-Frontier Experiments}},  \href{https://arxiv.org/abs/2207.00597}{{\ttfamily 2207.00597}}.

\bibitem{Ferber:2022ewf}
T.~Ferber, C.~Garcia-Cely and K.~Schmidt-Hoberg, \emph{{BelleII sensitivity to long{\textendash}lived dark photons}}, \href{https://doi.org/10.1016/j.physletb.2022.137373}{\emph{Phys. Lett. B} {\bfseries 833} (2022) 137373} [\href{https://arxiv.org/abs/2202.03452}{{\ttfamily 2202.03452}}].

\bibitem{LDMX:2018cma}
{\scshape LDMX} collaboration, \emph{{Light Dark Matter eXperiment (LDMX)}},  \href{https://arxiv.org/abs/1808.05219}{{\ttfamily 1808.05219}}.

\bibitem{LDMX:2025bog}
{\scshape LDMX} collaboration, \emph{{LDMX - The Light Dark Matter eXperiment}},  \href{https://arxiv.org/abs/2508.11833}{{\ttfamily 2508.11833}}.

\bibitem{Chang:2018rso}
J.H.~Chang, R.~Essig and S.D.~McDermott, \emph{{Supernova 1987A Constraints on Sub-GeV Dark Sectors, Millicharged Particles, the QCD Axion, and an Axion-like Particle}}, \href{https://doi.org/10.1007/JHEP09(2018)051}{\emph{JHEP} {\bfseries 09} (2018) 051} [\href{https://arxiv.org/abs/1803.00993}{{\ttfamily 1803.00993}}].

\bibitem{BaBar:2017tiz}
{\scshape BaBar} collaboration, \emph{{Search for Invisible Decays of a Dark Photon Produced in ${e}^{+}{e}^{-}$ Collisions at BaBar}}, \href{https://doi.org/10.1103/PhysRevLett.119.131804}{\emph{Phys. Rev. Lett.} {\bfseries 119} (2017) 131804} [\href{https://arxiv.org/abs/1702.03327}{{\ttfamily 1702.03327}}].

\bibitem{Andreev:2021fzd}
Y.M.~Andreev et~al., \emph{{Improved exclusion limit for light dark matter from e+e- annihilation in NA64}}, \href{https://doi.org/10.1103/PhysRevD.104.L091701}{\emph{Phys. Rev. D} {\bfseries 104} (2021) L091701} [\href{https://arxiv.org/abs/2108.04195}{{\ttfamily 2108.04195}}].

\bibitem{Banerjee:2019pds}
D.~Banerjee et~al., \emph{{Dark matter search in missing energy events with NA64}}, \href{https://doi.org/10.1103/PhysRevLett.123.121801}{\emph{Phys. Rev. Lett.} {\bfseries 123} (2019) 121801} [\href{https://arxiv.org/abs/1906.00176}{{\ttfamily 1906.00176}}].

\bibitem{COHERENT:2021pvd}
{\scshape COHERENT} collaboration, \emph{{First Probe of Sub-GeV Dark Matter beyond the Cosmological Expectation with the COHERENT CsI Detector at the SNS}}, \href{https://doi.org/10.1103/PhysRevLett.130.051803}{\emph{Phys. Rev. Lett.} {\bfseries 130} (2023) 051803} [\href{https://arxiv.org/abs/2110.11453}{{\ttfamily 2110.11453}}].

\bibitem{Baek:2011aa}
S.~Baek, P.~Ko and W.-I.~Park, \emph{{Search for the Higgs portal to a singlet fermionic dark matter at the LHC}}, \href{https://doi.org/10.1007/JHEP02(2012)047}{\emph{JHEP} {\bfseries 02} (2012) 047} [\href{https://arxiv.org/abs/1112.1847}{{\ttfamily 1112.1847}}].

\bibitem{Knapen:2017xzo}
S.~Knapen, T.~Lin and K.M.~Zurek, \emph{{Light Dark Matter: Models and Constraints}}, \href{https://doi.org/10.1103/PhysRevD.96.115021}{\emph{Phys. Rev. D} {\bfseries 96} (2017) 115021} [\href{https://arxiv.org/abs/1709.07882}{{\ttfamily 1709.07882}}].

\bibitem{BaBar:2013npw}
{\scshape BaBar} collaboration, \emph{{Search for $B \to K^{(*)} \nu \overline \nu$ and invisible quarkonium decays}}, \href{https://doi.org/10.1103/PhysRevD.87.112005}{\emph{Phys. Rev. D} {\bfseries 87} (2013) 112005} [\href{https://arxiv.org/abs/1303.7465}{{\ttfamily 1303.7465}}].

\bibitem{ATLAS:2015gvj}
{\scshape ATLAS} collaboration, \emph{{Search for invisible decays of a Higgs boson using vector-boson fusion in $pp$ collisions at $\sqrt{s}=8$ TeV with the ATLAS detector}}, \href{https://doi.org/10.1007/JHEP01(2016)172}{\emph{JHEP} {\bfseries 01} (2016) 172} [\href{https://arxiv.org/abs/1508.07869}{{\ttfamily 1508.07869}}].

\bibitem{CRESST:2015txj}
{\scshape CRESST} collaboration, \emph{{Results on light dark matter particles with a low-threshold CRESST-II detector}}, \href{https://doi.org/10.1140/epjc/s10052-016-3877-3}{\emph{Eur. Phys. J. C} {\bfseries 76} (2016) 25} [\href{https://arxiv.org/abs/1509.01515}{{\ttfamily 1509.01515}}].

\bibitem{Krnjaic:2015mbs}
G.~Krnjaic, \emph{{Probing Light Thermal Dark-Matter With a Higgs Portal Mediator}}, \href{https://doi.org/10.1103/PhysRevD.94.073009}{\emph{Phys. Rev. D} {\bfseries 94} (2016) 073009} [\href{https://arxiv.org/abs/1512.04119}{{\ttfamily 1512.04119}}].

\bibitem{Nguyen:2025tkl}
T.T.Q.~Nguyen, P.~De~la Torre~Luque, I.~John, S.~Balaji, P.~Carenza and T.~Linden, \emph{{INTEGRAL, eROSITA and Voyager constraints on light bosonic dark matter: ALPs, dark photons, scalars, B-L and Li-Lj vectors}}, \href{https://doi.org/10.1103/fn6k-1nlc}{\emph{Phys. Rev. D} {\bfseries 113} (2026) 103010} [\href{https://arxiv.org/abs/2507.13432}{{\ttfamily 2507.13432}}].

\bibitem{DelaTorreLuque:2025zjt}
P.~De~la Torre~Luque, P.~Carenza and T.T.Q.~Nguyen, \emph{{Sub-keV dark matter can strongly ionize molecular clouds}}, \href{https://doi.org/10.1103/qmw9-1k4k}{\emph{Phys. Rev. D} {\bfseries 113} (2026) 063035} [\href{https://arxiv.org/abs/2507.01962}{{\ttfamily 2507.01962}}].

\bibitem{Balaji:2025alr}
S.~Balaji, P.~Carenza, P.~De~la Torre~Luque, A.~Lella and L.~Mastrototaro, \emph{{In-flight positron annihilation as a probe of feebly interacting particles}}, \href{https://doi.org/10.1103/PhysRevD.111.083053}{\emph{Phys. Rev. D} {\bfseries 111} (2025) 083053} [\href{https://arxiv.org/abs/2501.07725}{{\ttfamily 2501.07725}}].

\bibitem{Oka:2005}
T.~Oka, T.R.~Geballe, M.~Goto, T.~Usuda and B.J.~McCall, \emph{{H$_3^+$ Spectroscopy and the Ionization Rate of Molecular Hydrogen in the Central Few Parsecs of the Galaxy}}, \href{https://doi.org/10.1086/432679}{\emph{The Astrophysical Journal} {\bfseries 632} (2005) 882}.

\bibitem{Ravikularaman_2025}
S.~Ravikularaman, S.~Recchia, V.H.M.~Phan and S.~Gabici, \emph{Cosmic rays cannot explain the high ionisation rates in the galactic centre}, \href{https://doi.org/10.1051/0004-6361/202451155}{\emph{Astronomy \& Astrophysics} {\bfseries 694} (2025) A114}.

\bibitem{DelaTorreLuque:2024fcc}
P.~De~la Torre~Luque, S.~Balaji and J.~Silk, \emph{{Anomalous Ionization in the Central Molecular Zone by Sub-GeV Dark Matter}}, \href{https://doi.org/10.1103/PhysRevLett.134.101001}{\emph{Phys. Rev. Lett.} {\bfseries 134} (2025) 101001} [\href{https://arxiv.org/abs/2409.07515}{{\ttfamily 2409.07515}}].

\bibitem{porter2008inverse}
T.A.~Porter, I.V.~Moskalenko, A.W.~Strong, E.~Orlando and L.~Bouchet, \emph{Inverse compton origin of the hard x-ray and soft gamma-ray emission from the galactic ridge}, {\emph{The Astrophysical Journal} {\bfseries 682} (2008) 400}.

\bibitem{Orlando2018MNRAS.475.2724O}
E.~{Orlando}, \emph{{Imprints of cosmic rays in multifrequency observations of the interstellar emission}}, \href{https://doi.org/10.1093/mnras/stx3280}{\emph{MNRAS} {\bfseries 475} (2018) 2724} [\href{https://arxiv.org/abs/1712.07127}{{\ttfamily 1712.07127}}].

\bibitem{tomsick2019comptonspectrometerimager}
J.A.~Tomsick, A.~Zoglauer, C.~Sleator, H.~Lazar, J.~Beechert, S.~Boggs et~al., \emph{The compton spectrometer and imager},  2019.

\end{thebibliography}\endgroup

\clearpage
\appendix

\section{Systematic uncertainties}
\label{App}

\begin{figure}[h]
    \centering
    \includegraphics[width=0.49\linewidth]{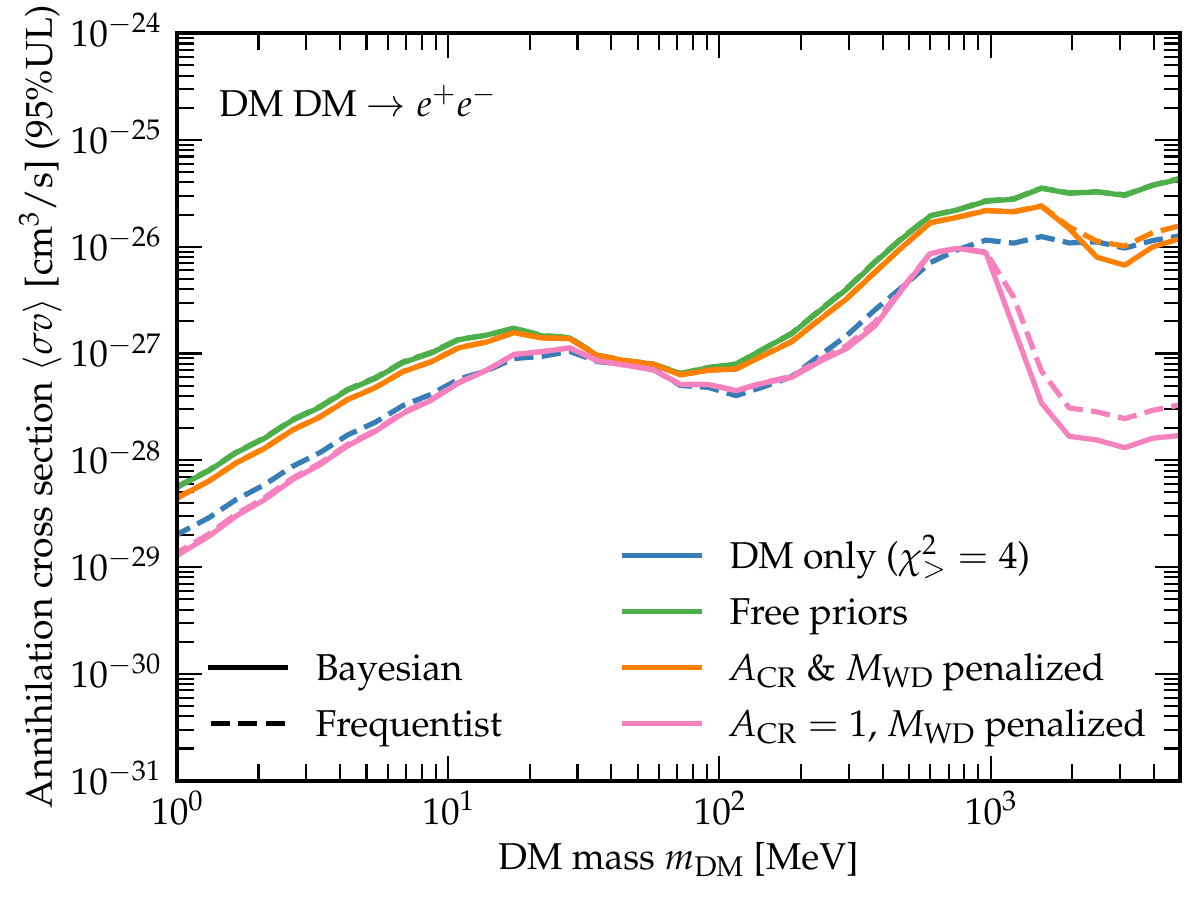}
    \includegraphics[width=0.49\linewidth]{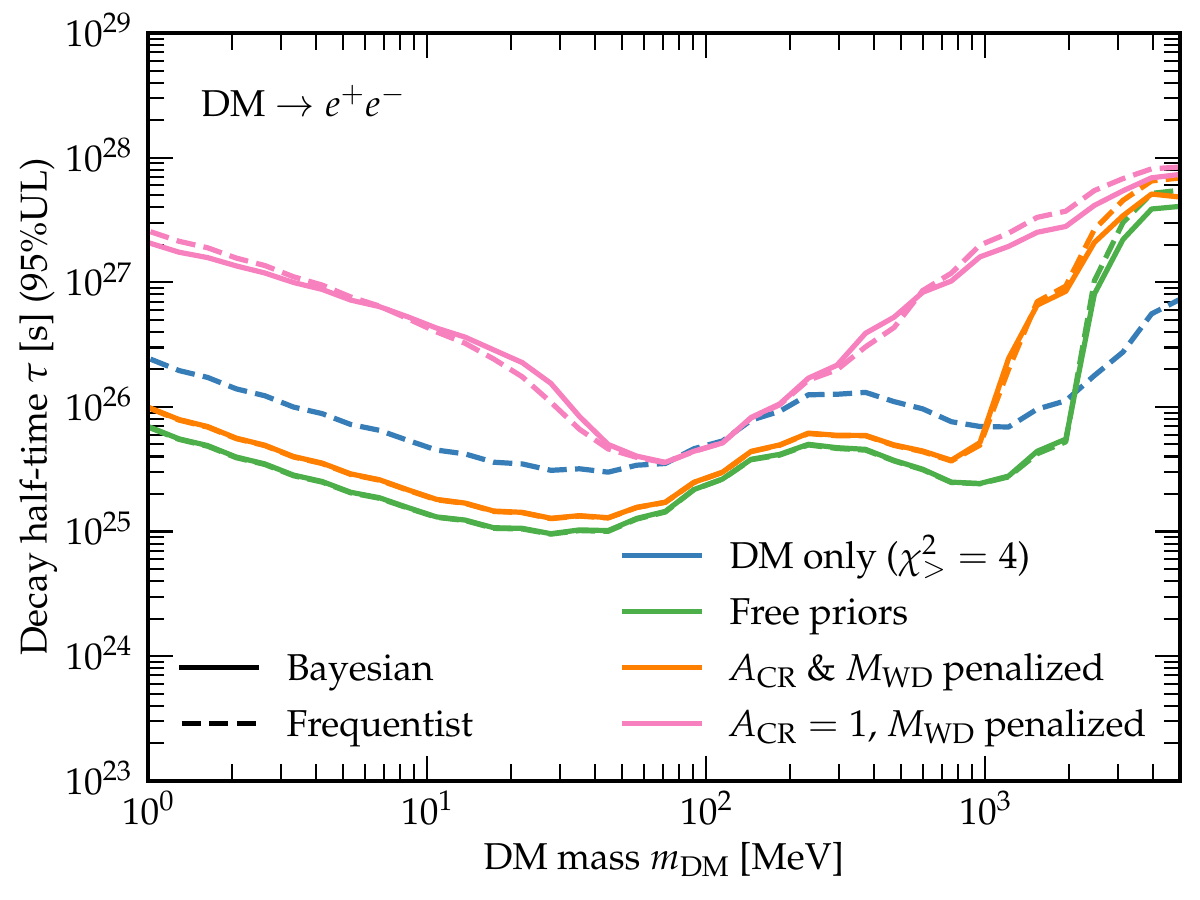}
    \caption{Impact of the prior scenario and statistical method on the derived constraints for DM annihilation into $e^+e^-$ (left) and decay (right). Solid lines correspond to the Bayesian MCMC analysis and dashed lines to the frequentist profile-likelihood approach. The three prior scenarios are shown: free priors (green), $A_{\rm CR}\,\&\,M_{\rm WD}$ penalized (orange, benchmark), and $A_{\rm CR}=1$ with $M_{\rm WD}$ penalized (pink). The conservative DM-only constraint ($\chi^2_> = 4$, blue) is also shown. The spread between the free-prior and penalized curves illustrates the systematic uncertainty associated with the astrophysical background modeling.}
    \label{fig:unc_priors}
\end{figure}

\begin{figure}[h]
    \centering
    \includegraphics[width=0.49\linewidth]{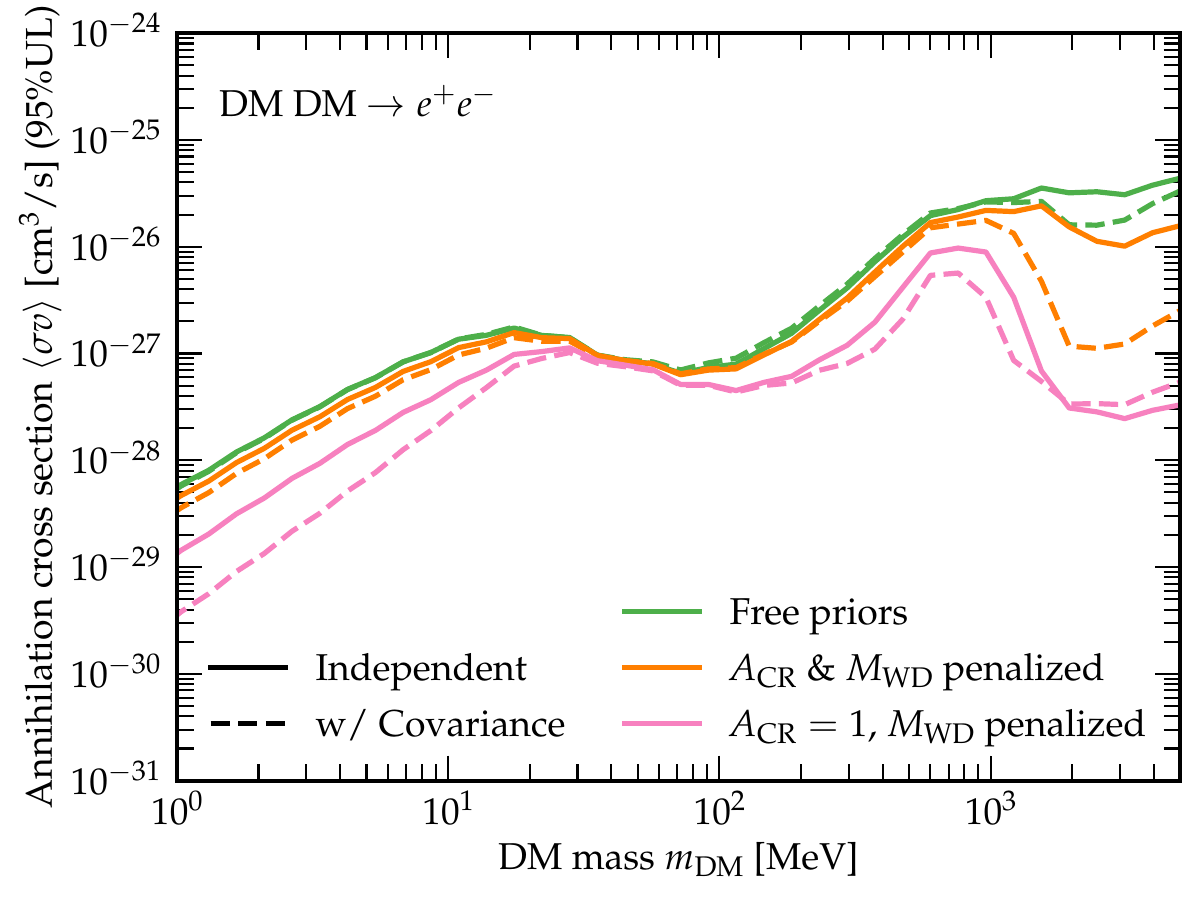}
    \includegraphics[width=0.49\linewidth]{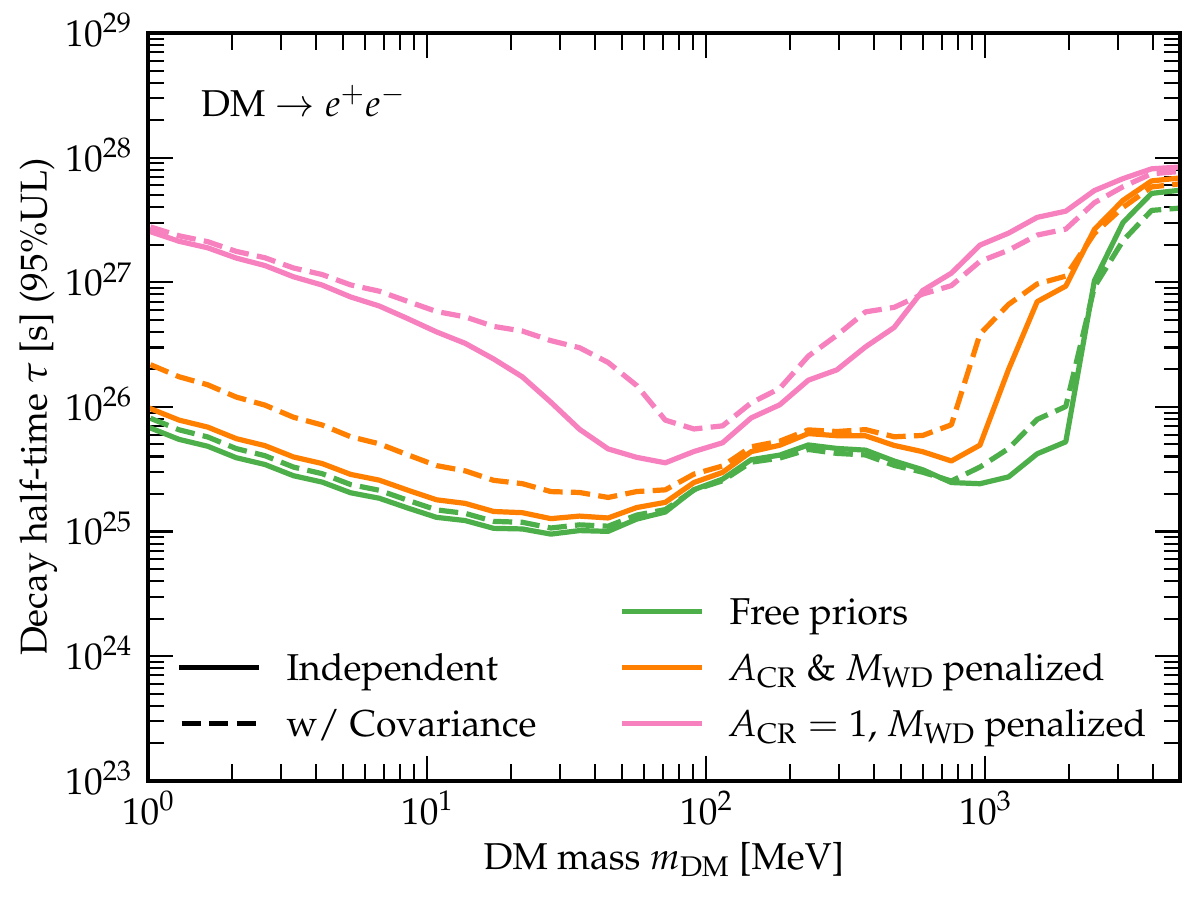}
    \caption{Impact of accounting for correlations between measurements on the derived constraints for DM annihilation into $e^+e^-$ (left) and decay (right), for the three prior scenarios. Solid lines show constraints obtained without the covariance matrix (independent measurements); dashed lines include the full estimated covariance. The covariance has a modest effect on the resulting limits, generally yielding slightly more conservative constraints.}
    \label{fig:unc_cov}
\end{figure}

\begin{figure}[t]
    \centering
    \includegraphics[width=0.49\linewidth]{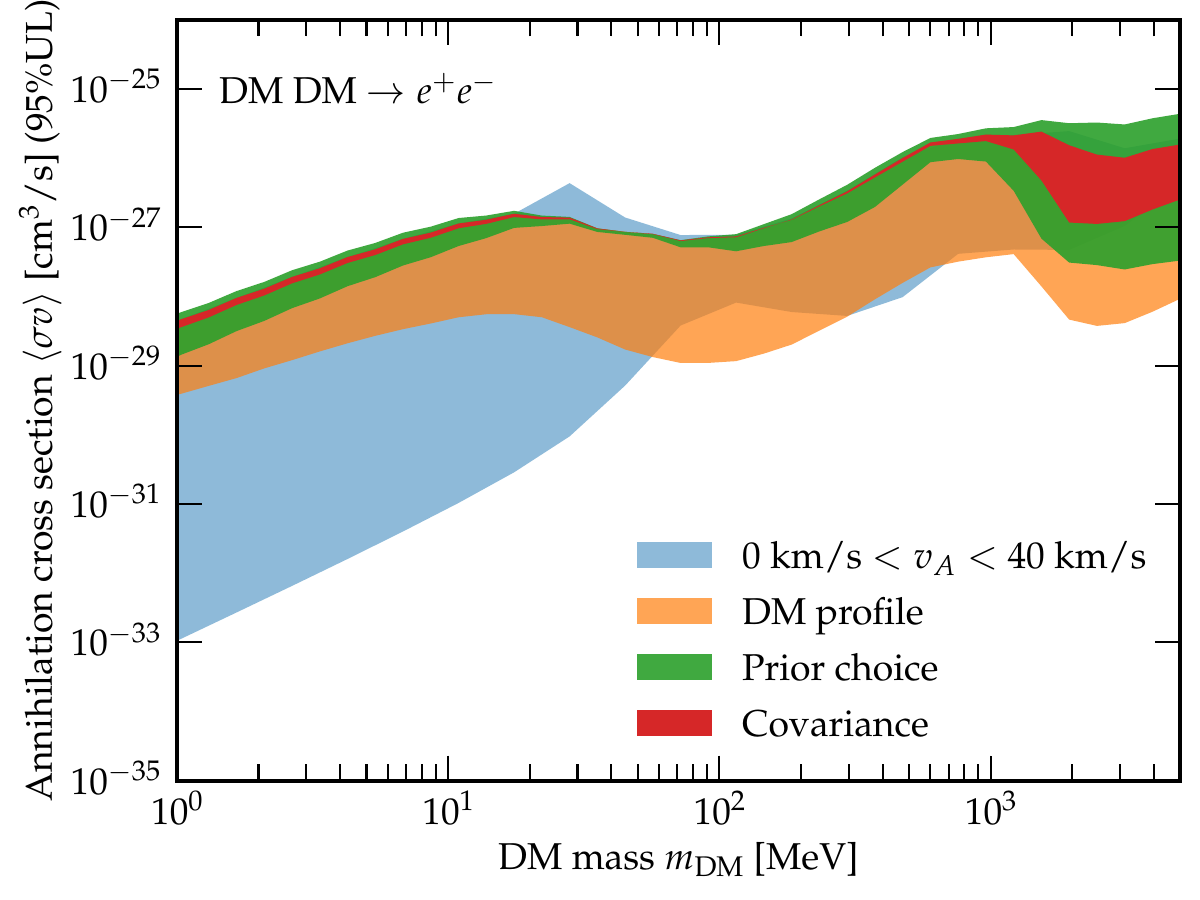}
    \includegraphics[width=0.49\linewidth]{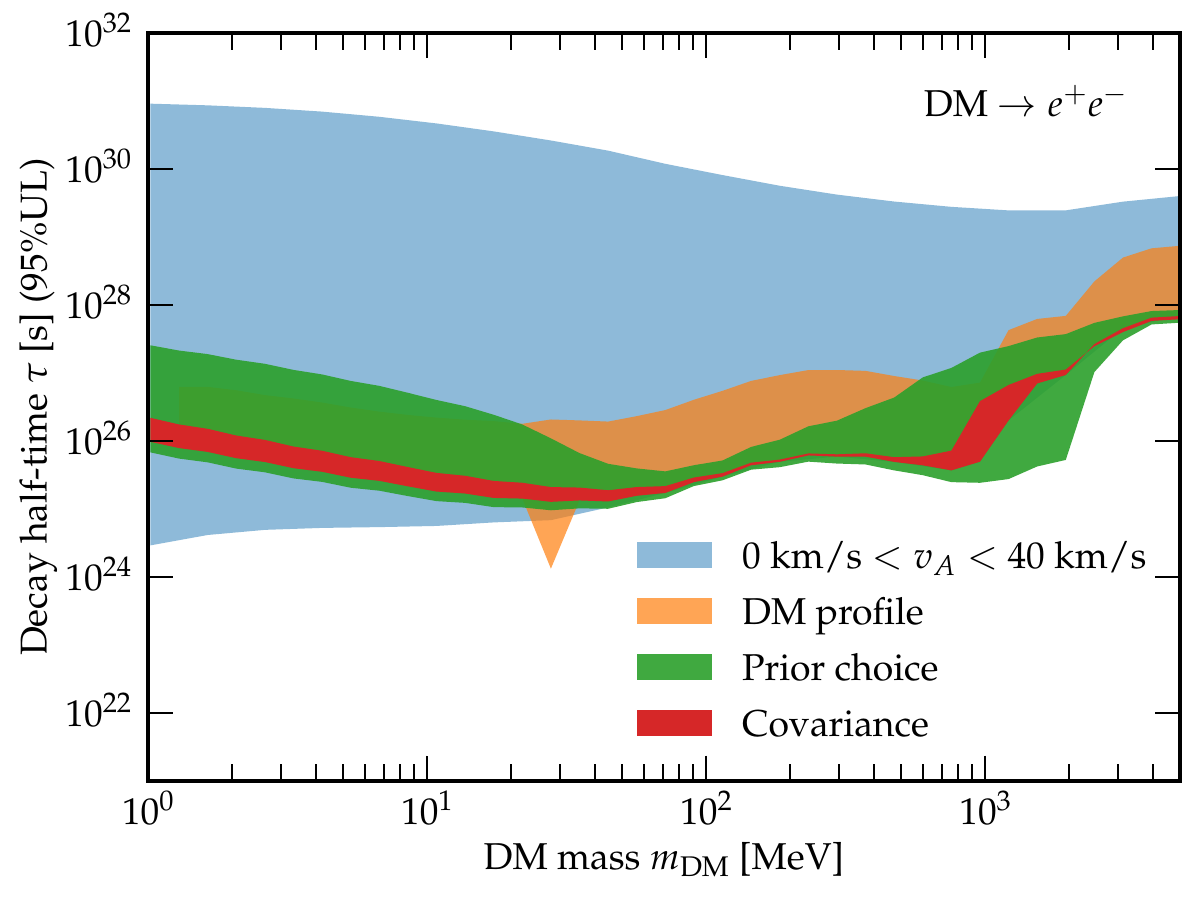}
    \caption{Systematic uncertainty bands on the benchmark constraints for DM annihilation into $e^+e^-$ (left) and decay (right). The shaded regions show the variation induced by the main sources of uncertainty: CR propagation parametrized by the Alfvén velocity $v_A \in [0, 40]$\,km\,s$^{-1}$ (blue), DM density profile (orange, from isothermal to generalized NFW with $\gamma=1.5$), prior scenario (green), and covariance treatment (red).}
    \label{fig:unc_prop}
\end{figure}

In this appendix, we assess the robustness of our benchmark constraints against the main sources of systematic uncertainty: the choice of prior scenario for the nuisance parameters, the treatment of correlations between measurements, the CR propagation model, and the assumed DM density profile. The results are mainly summarised in Figures~\ref{fig:unc_priors} to~\ref{fig:unc_prop}.

\subsection{Prior scenario and statistical method}

Figure~\ref{fig:unc_priors} shows the impact of the prior scenario and statistical method on the derived constraints for the $e^+e^-$ final state. The Bayesian and frequentist analyses yield very similar results across the full mass range, confirming that the posterior distributions are approximately Gaussian for the DM signal parameter. The spread between the free-prior and penalized scenarios provides a direct estimate of the systematic uncertainty associated with the astrophysical background modeling: constraints derived with free priors are generally weaker, since the astrophysical components have more freedom to absorb a putative DM signal. The $A_{\rm CR}=1$, $M_{\rm WD}$ penalized case yields the tightest limits, as fixing the CR normalization removes a degree of freedom that is partially degenerate with the DM signal at high energies. The conservative DM-only constraint ($\chi^2_>=4$) closely tracks the free-prior curve, confirming that the free-prior scenario provides a conservative upper envelope even when including backgrounds. Since the statistical test between the conservative DM-only analysis and the analysis with the inclusion of the astrophysical background ($\chi^2=2.71$) is different, it is not expected that the DM-only case always provides weaker constraints.

\subsection{Measurement correlations}

Figure~\ref{fig:unc_cov} illustrates the impact of including the estimated covariance matrix in the analysis. Accounting for correlations between measurements — arising from shared INTEGRAL revolutions and a common background-modeling procedure — has a modest effect on the derived constraints, typically strengthening them by a factor of a few at most. This is consistent with the finding in Ref.~\cite{Koechler:2026dci} for the PBH case. Given that the covariance treatment represents only an estimation, to remain conservative we adopt the treatment without the covariance matrix as our benchmark.

\subsection{CR propagation and DM profile}

Figure~\ref{fig:unc_prop} shows the total systematic uncertainty band for the $e^+e^-$ channel, decomposed by source. The dominant uncertainty arises from CR propagation, specifically from the Alfvén velocity $v_A$ controlling diffusive reacceleration. Varying $v_A$ from 0 to 40 km/s spans the full blue band, which dominates all other sources of uncertainty across the mass range. This is consistent with findings from previous analyses~\cite{DelaTorreLuque:2023olp, Balaji:2025afr}: reacceleration shifts the energy of the propagated electrons, modifying both the spectral shape and the spatial extent of the IC signal, with strong reacceleration producing a harder and more extended emission. The direct impact of reacceleration in the morphology and spectrum of the signals is clearly observed in Figure~\ref{fig:flux_vA}.

The DM density profile uncertainty (orange band) is comparatively small, reflecting the fact that the DM-induced signal scales linearly with $\rho_{\rm DM}$ for decay and quadratically for annihilation, but the IC emission is spatially smoothed by propagation over several degrees. Variations in the profile between isothermal and cuspy NFW shapes therefore have a limited impact on the predicted signal in the sky regions and energy bands used in this analysis. The prior choice and covariance treatment (green and red bands, respectively) introduce the smallest variations, confirming the robustness of our benchmark results against these aspects of the analysis.

\subsection{Nuisance parameter behaviour}

\begin{figure}[t]
    \centering
    \includegraphics[width=0.49\linewidth]{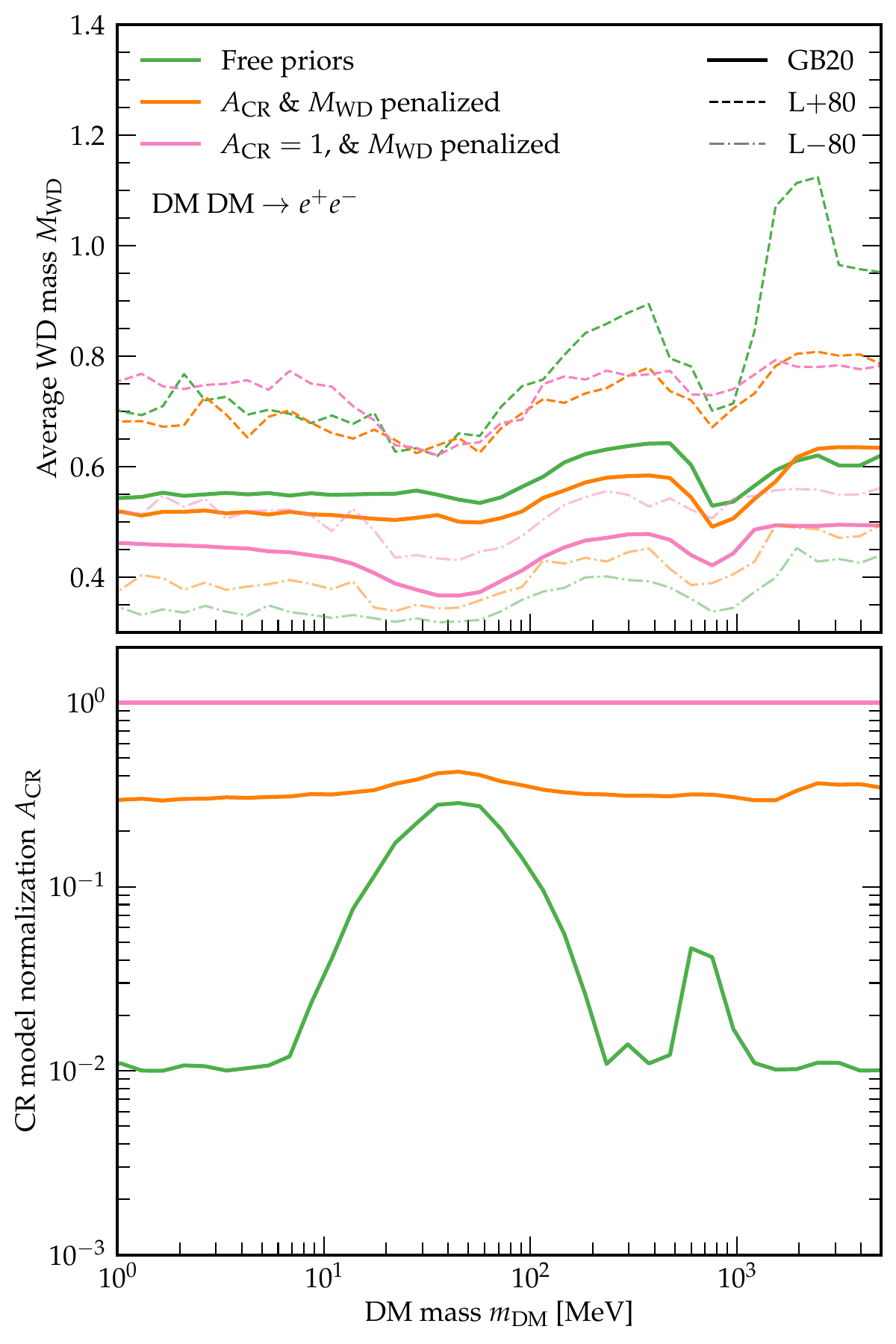}
    \includegraphics[width=0.49\linewidth]{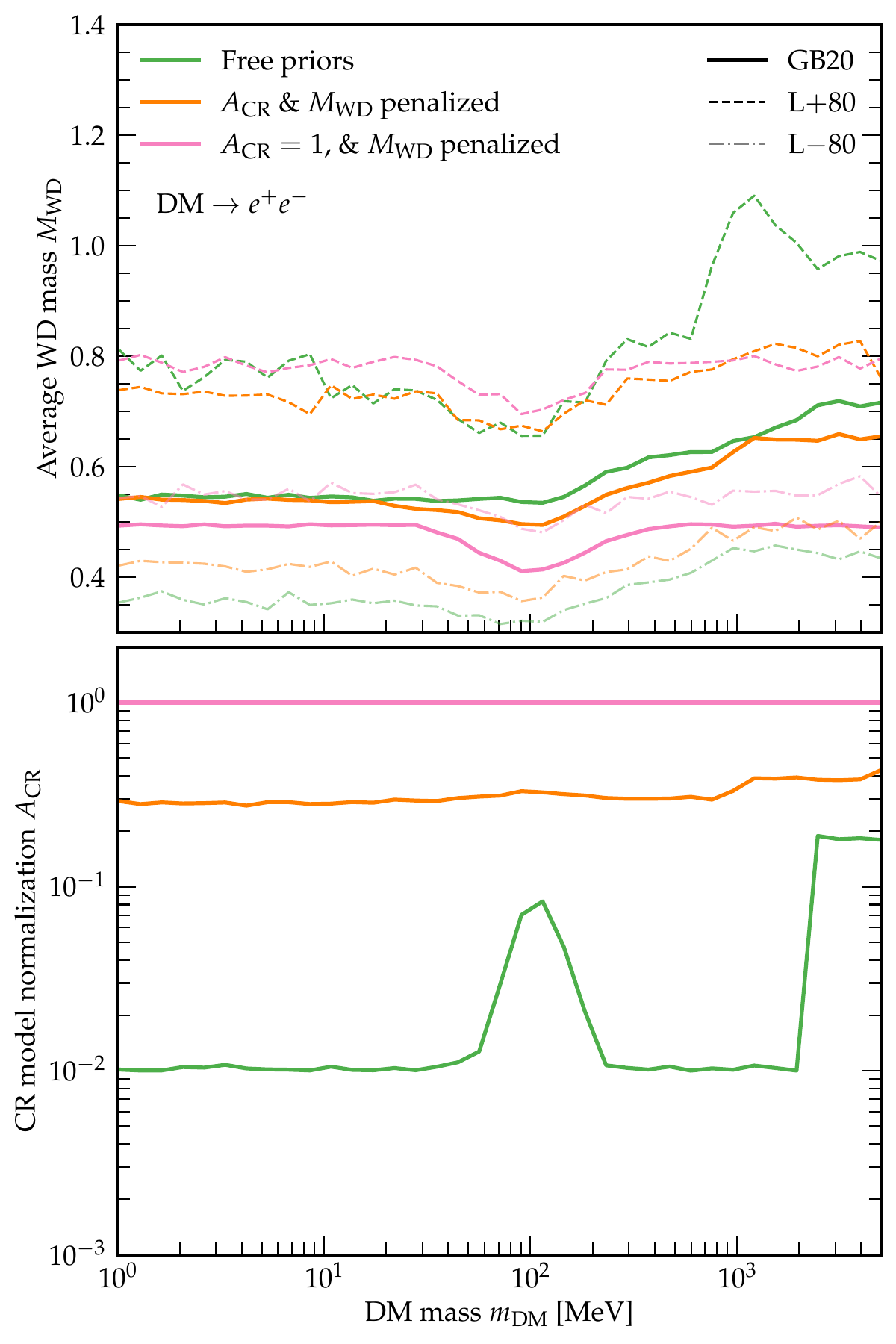}
    \caption{Best-fit values of the nuisance parameters as a function of DM mass $m_{\rm DM}$ for the $e^+e^-$ final state, for DM annihilation (left) and decay (right). Upper panels: average WD masses $M_{\rm WD}$ in the three sky regions (\texttt{GB20}, solid; \texttt{L+80}, dashed; \texttt{L-80}, dot-dashed), for the three prior scenarios (green: free priors; orange: $A_{\rm CR}\,\&\,M_{\rm WD}$ penalized; pink: $A_{\rm CR}=1$, $M_{\rm WD}$ penalized). Lower panels: best-fit CR normalization $A_{\rm CR}$. In the free-prior case, $A_{\rm CR}$ is driven significantly below unity over the mass range where the DM signal prefers to account for the high-energy Galactic centre emission, illustrating the partial degeneracy between the CR and DM components discussed in Section~\ref{sec:results}.}
    \label{fig:fitDM_MWDACR}
\end{figure}

Figure~\ref{fig:fitDM_MWDACR} shows the best-fit nuisance parameter values as a function of DM mass for the $e^+e^-$ final state. The WD masses remain in the range $0.4$--$0.8\,M_\odot$ across all regions and prior scenarios, in good agreement with previous measurements~\cite{Krivonos:2006px, Turler:2010pm, Yuasa:2012qe}. The \texttt{L+80} region consistently prefers slightly higher WD masses than \texttt{GB20} and \texttt{L-80}, reflecting possible spatial variations in the WD population between the inner and outer Galaxy.

The CR normalization $A_{\rm CR}$ shows a more complex behaviour. In the penalized scenarios it remains being a fraction to unity over all these mass range. In the free-prior case, however, $A_{\rm CR}$ is driven substantially below unity over the mass range corresponding to the $\Delta\chi^2$ preference discussed in Section~\ref{sec:results}, as the fit redistributes high-energy Galactic center emission from the CR component to the DM signal. This behaviour is more pronounced for annihilation than for decay, consistent with the stronger central concentration of the annihilation signal. The penalized prior on $A_{\rm CR}$ suppresses this degeneracy and yields more physically motivated background parameters, which is the primary motivation for adopting it as our benchmark.

\subsection{Impact of CR propagation on the predicted signal}

\begin{figure}[t]
    \centering
    \includegraphics[width=\linewidth]{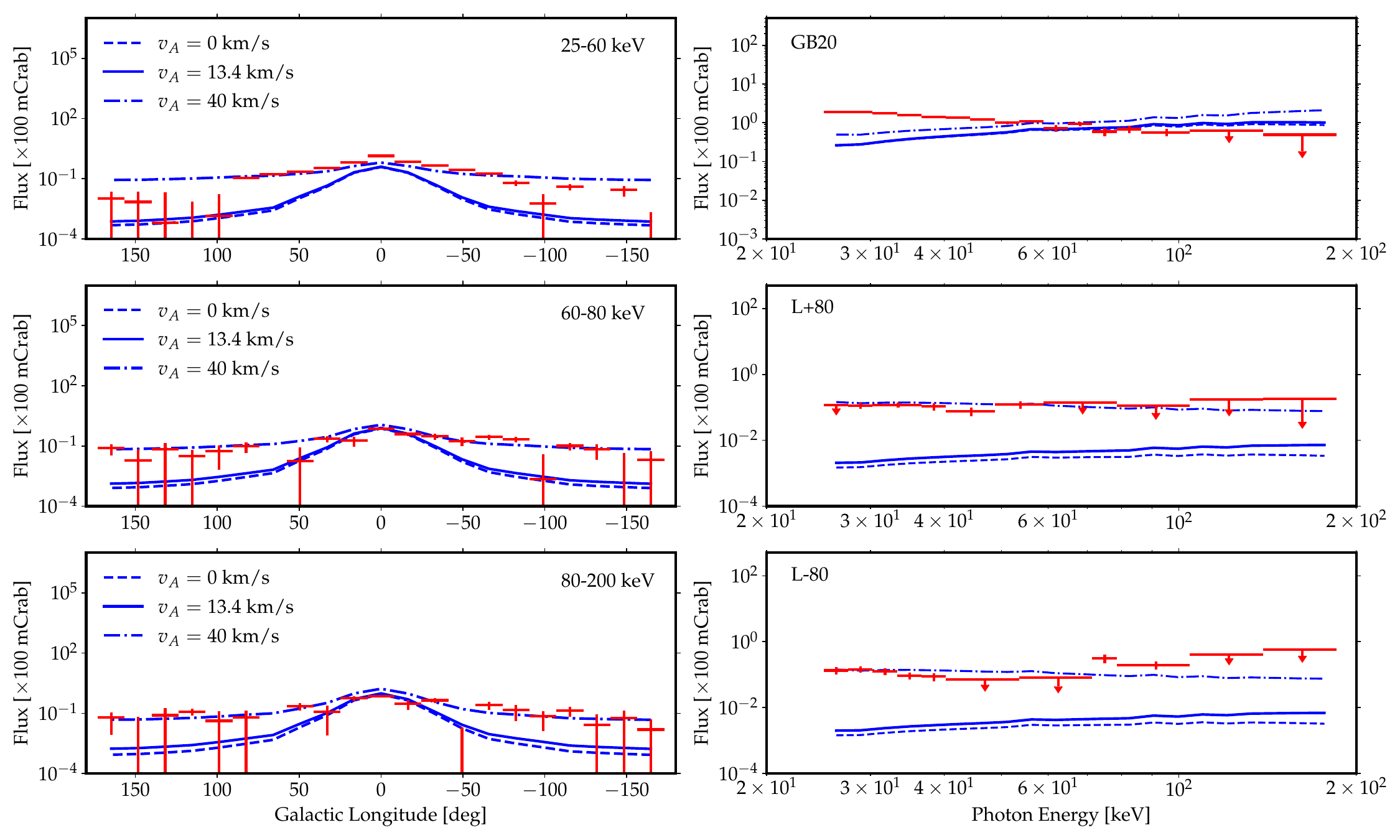}
    \caption{Morphology and spectra of the DM annihilation signal into $e^+e^-$ for $m_{\rm DM} = 100$\,MeV, under three CR propagation scenarios: benchmark ($v_A = 13.4$ km/s, solid blue), no reacceleration ($v_A = 0$ km/s, dashed blue), and strong reacceleration ($v_A = 40$ km/s, dot-dashed blue). The IBIS/ISGRI measurements are shown as red points. Strong reacceleration produces a harder and more spatially extended IC signal, as electrons are boosted to higher energies and propagate farther before losing energy, while the absence of reacceleration yields a softer, more concentrated morphology.}
    \label{fig:flux_vA}
\end{figure}

Figure~\ref{fig:flux_vA} illustrates how the predicted DM signal morphology and spectrum depend on the CR propagation model through the choice of $v_A$. Strong reacceleration ($v_A = 40$ km/s) shifts the IC emission to higher energies and broadens the spatial morphology, as the injected electrons gain energy through wave--particle interactions and propagate farther before cooling. In the absence of reacceleration ($v_A=0$), the electron spectrum becomes softer and the spatial distribution more centrally concentrated. This is because, without reacceleration, electrons remain being less energetic and lose energy more efficiently, therefore having shorter diffusion lengths, limiting their propagation away from their sources. These differences directly translate into the large propagation uncertainty band seen in Figure~\ref{fig:unc_prop}.

\clearpage

\section{Additional plots}
\label{app:additional}

\begin{figure}[hb]
    \centering
    \includegraphics[width=\linewidth]{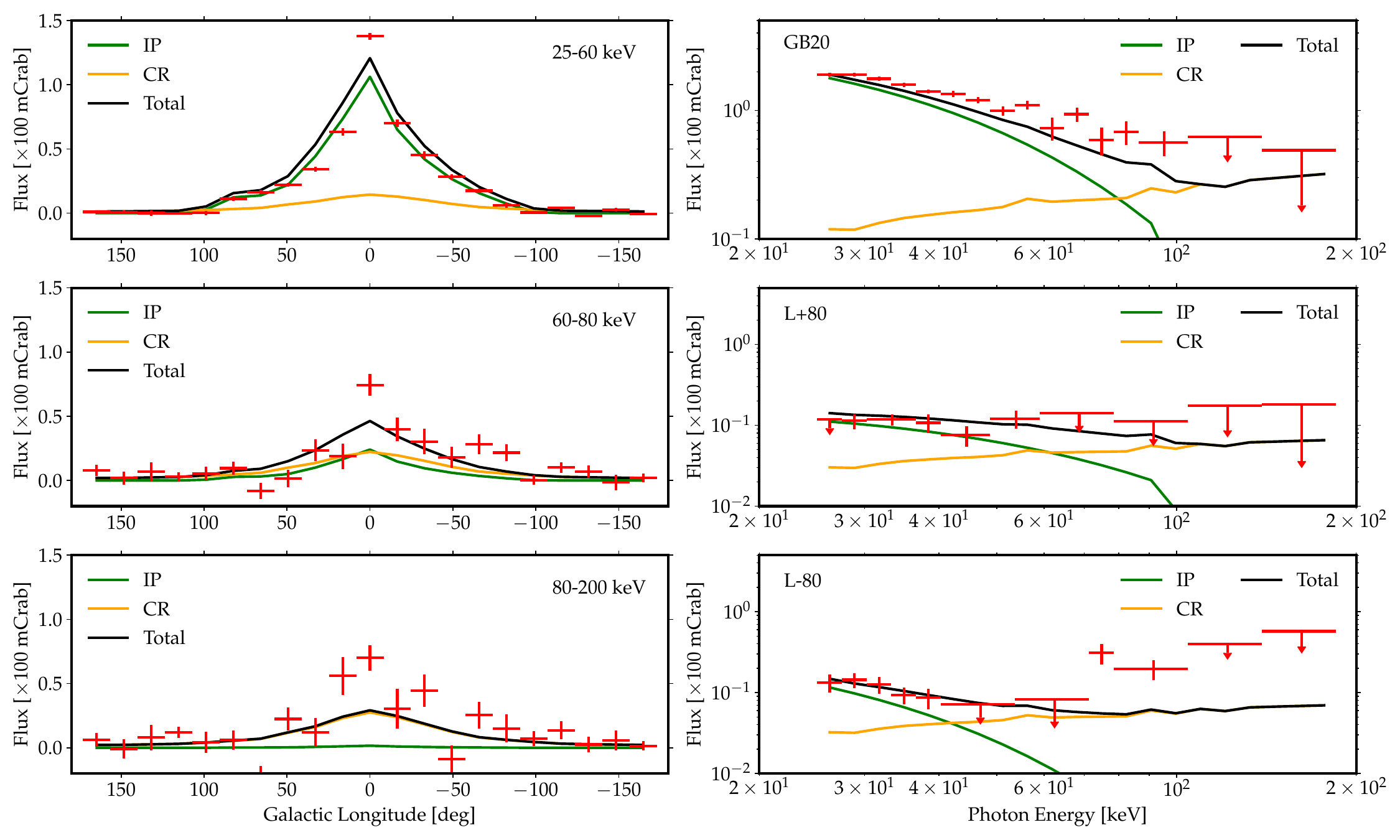}
    \caption{Morphology and spectra of the X-ray emissions, the red points being the measurements from IBIS/ISGRI. The benchmark best-fit CR contribution is in orange, the best-fit IP contribution from WDs in green and the sum of all components is shown in black. This plot assumes the $A_{\rm CR}\,\&\,M_{\rm WD}$ penalized prior scenario, with the following fitted parameters: $A_{\rm CR}=0.39$, $A_{\rm IP}^\texttt{GB20}=4.30\times10^{-6}$ and $\{M_{\rm WD}^\texttt{GB20},M_{\rm WD}^\texttt{L+80},M_{\rm WD}^\texttt{L-80}\}=\{0.65,0.82,0.48\}M_\odot$. The associated best-fit chi-square is $\chi^2_{\rm eff}/\rm{d.o.f.}=5.48$.}
    \label{fig:fluxnoDMBF}
\end{figure}

We report here additional figures for completeness. Figure~\ref{fig:fluxnoDMBF} shows the best-fit background-only morphology and energy spectra — without any DM contribution — in the three sky regions and three energy bands, obtained under the $A_{\rm CR}\,\&\,M_{\rm WD}$ penalized prior scenario (our benchmark). This illustrates the quality of the two-component astrophysical fit to the IBIS data, with the WD (IP) component dominating at low energies and the CR component accounting for the high-energy tail. 

\begin{figure}[t]
    \centering
    \includegraphics[width=\linewidth]{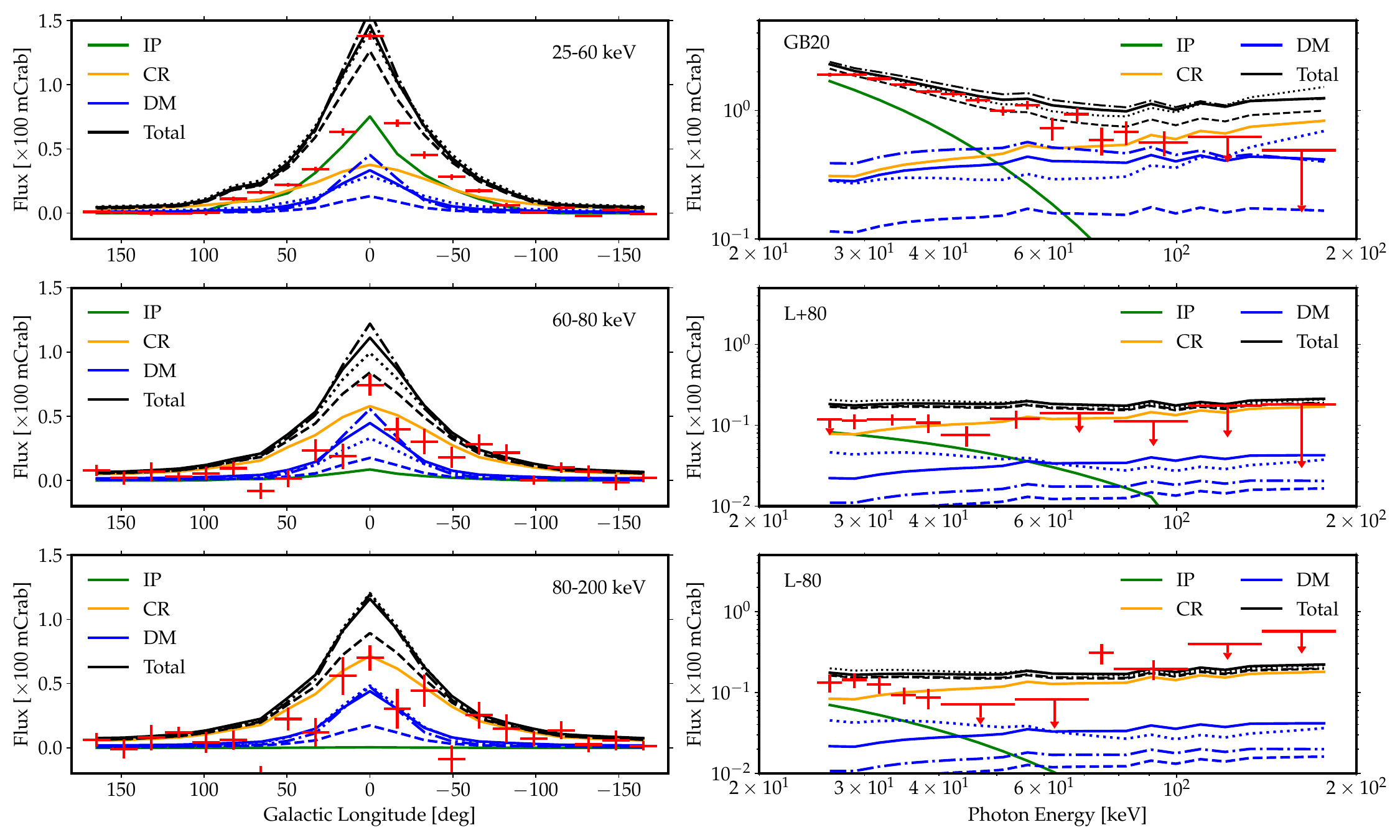}
    \caption{Morphology and spectra of the X-ray emissions, the red points being the measurements from IBIS/ISGRI. The DM-induced emissions from the decay into $e^+e^-$ (in blue) are from four cases: $m_{\rm DM} =1$ MeV and $\tau^{-1}=10^{26}$ s$^{-1}$ (solid), $m_{\rm DM} =10$ MeV and $\tau^{-1}=2\times10^{26}$ s$^{-1}$ (dashed), $m_{\rm DM} =100$ MeV and $\tau^{-1}=3\times10^{26}$ s$^{-1}$ (dot-dashed), $m_{\rm DM} =1$ GeV and $\tau^{-1}=4\times10^{26}$ s$^{-1}$ (dotted). The predicted CR contribution is in orange, the IP contribution from WDs in green and the sum of all components is shown in black.}
    \label{fig:fluxmDMdec}
\end{figure}

Figure~\ref{fig:fluxmDMdec} shows the equivalent of Figure~\ref{fig:fluxmDMann} in the main text for the case of DM decay into $e^+e^-$, displaying the predicted morphology and spectra for four representative DM masses and decay rates. As discussed in Section~\ref{sec:results}, the decay signal is morphologically more extended and less centrally concentrated than the annihilation case, reflecting the linear rather than quadratic dependence on the DM density profile.

\end{document}